\documentclass[twocolumn]{aastex63}

\hypersetup{linkcolor=red,citecolor=blue,filecolor=cyan,urlcolor=magenta}

\usepackage[]{units}
\usepackage{gensymb}
\usepackage{amsmath}
\usepackage{booktabs}
\usepackage{romannum}
\usepackage{longtable}
\usepackage{xspace}
\usepackage{enumitem}

\newcommand{\ha}{H\ensuremath{\alpha}}

\shorttitle{UHR Signature}
\shortauthors{Krishnarao et al.}

\graphicspath{{./}{figures_png/}}

\begin{document}

\title{Photometric Signature of Ultra-Harmonic Resonances in Barred Galaxies}

\correspondingauthor{Dhanesh Krishnarao}
\email{dkrishnarao@coloradocollege.edu}

\author[0000-0002-7955-7359]{Dhanesh Krishnarao}
\affiliation{Department of Astronomy, University of Wisconsin-Madison, Madison, WI, USA}
\affiliation{NSF Astronomy \& Astrophysics Postdoctoral Fellow, Johns Hopkins University, Baltimore, MD, USA}
\affiliation{Department of Physics, Colorado College, Colorado Springs, CO, USA}

\author[0000-0003-4843-4185]{Zachary J. Pace}
\affiliation{Department of Astronomy, University of Wisconsin-Madison, Madison, WI, USA}

\author{Elena D'Onghia}
\affiliation{Department of Astronomy, University of Wisconsin-Madison, Madison, WI, USA}

\author{J. Alfonso L. Aguerri}
\affiliation{Instituto de Astrofísica de Canarias, Tenerife, Spain}
\affiliation{Departamento de Astrofísica, Universidad de La Laguna, Tenerife, Spain}

\author[0000-0001-5928-7155]{Rachel L. McClure}
\affiliation{Department of Astronomy, University of Wisconsin-Madison, Madison, WI, USA}

\author[0000-0003-3217-7778]{Thomas Peterken}
\affiliation{School of Physics and Astronomy, University of Nottingham, University Park, Nottingham, UK}

\author[0000-0003-3526-5052]{Jos\'e G. Fern\'andez-Trincado}
\affiliation{Instituto de Astronom\'ia, Universidad Cat\'olica del Norte, Av. Angamos 0610, Antofagasta, Chile}
\affiliation{Universidad de Atacama, Copayapu 485, Copiap\'o, Chile}

\author[0000-0002-4202-4727]{Michael Merrifield}
\affil{School of Physics and Astronomy, University of Nottingham, University Park, Nottingham, UK}

\author[0000-0003-0846-9578]{Karen L. {Masters}}
\affil{Department of Physics \& Astronomy, Haverford College, Haverford, PA, USA}

\author{Luis Garma-Oehmichen}
\affil{Instituto de Astronomía, Universidad Nacional Autónoma de México, Apartado Postal 70-264, México D.F., 04510, México}

\author{Nicholas Fraser Boardman}
\affiliation{Department of Physics and Astronomy, University of Utah, Salt Lake City, UT, USA}

\author[0000-0002-3131-4374]{Matthew Bershady}
\affiliation{Department of Astronomy, University of Wisconsin-Madison, Madison, WI, USA}
\affiliation{South African Astronomical Observatory, Cape Town, South Africa}
\affiliation{Department of Astronomy, University of Cape Town, Private Bag X3, Rondebosch 7701, South Africa}

\author[0000-0002-7339-3170]{Niv {Drory}}
\affiliation{McDonald Observatory, The University of Texas at Austin, Austin, TX, USA}

\author{Richard R. Lane}
\affiliation{Instituto de Astronomi\'a y Ciencias Planetarias de Atacama, Universidad de Atacama, Copayapu 485, Copiapó, Chile}

\begin{abstract}

Bars may induce morphological features, such as rings, through their resonances. Previous studies suggested that the presence of `dark-gaps', or regions of a galaxy where the difference between the surface brightness along the bar major axis and along the bar minor axis are maximal, can be attributed to the location of bar corotation.
Here, using GALAKOS, a high-resolution N-body simulation of a barred galaxy, we test this photometric method's ability to identify the bar corotation resonance.
Contrary to previous work, our results indicate that `dark-gaps' are a clear sign of the location of the 4:1 ultra-harmonic resonance instead of bar corotation.  Measurements of the bar corotation
can indirectly be inferred using kinematic information, e.g., by measuring the shape of the rotation curve. 
We demonstrate our concept on a sample of $578$ face-on barred galaxies with both imaging and integral field observations and find the sample likely consists primarily of fast bars. 
\end{abstract}


\section{Introduction} \label{sec:intro}

Observations reveal that $50\%$ to over $70\%$ of nearby disk galaxies host a bar \citep[e.g.][]{Eskridge2000, Aguerri2009,Nair2010, Masters2011} including our own Milky Way \citep{Binney1991,Weiland1994, Hammersley2001,Benjamin2005}. 
Bars are long-lived features with significant effects on the stellar and gas distribution and kinematics throughout galaxies \citep{Athanassoula2005,Barazza2008,Sheth2008}.
To inform a complete picture of galaxy evolution, we must understand the formation of bars and their impact on a galaxy's dynamics and evolutionary track.
Bars form both as a result of interactions \citep{Noguchi1987,Elmegreen1991,Romano-Diaz2008,Martinez-Valpuesta2016} and spontaneously from gravitational instabilities \citep{Toomre1964,Polyachenko2013}.
The strong non-axisymmetric nature of bars allows for the transport of angular momentum, energy, and mass across large radial scales within a galaxy \citep[e.g.][]{Debattista2000, Athanassoula2003}. 

Additionally, bars are thought to rotate as a solid body with an angular velocity called the pattern speed ($\Omega_p$) and can thus drive resonances with the rotating matter in the disk beyond the physical extent of the bar itself. 
The pattern speed of bars is one of the fundamental parameters necessary to understand the dynamics of disk galaxies. 
Corotation is defined as the radius at which the galaxy's circular angular rotation is equal to the bar pattern speed.
The bar pattern speed also often determines the locations of the 2:1 Lindblad resonance \citep{Lindblad1941} and the 4:1 ultra-harmonic resonance, which can both induce the formation of structures, such as rings \citep[e.g.][]{Buta1986}.

These dynamical parameters and bar resonance locations typically require stellar kinematics measurements to derive, with \citet{Tremaine1984} outlining the most accurate and only model-independent method. 
Some attempts to tie morphological features, such as rings, to the locations of resonances have been made \citep[e.g.][]{Buta1986, Buta2017}. 
The locations of such features could then be used to infer resonance locations and pattern speeds, but are model-dependent. 
In particular, \citet{Buta2017} proposed that the location of `dark-gaps', or regions of a galaxy along the bar minor axis that are dimmer in surface brightness than the equivalent radius along the bar major axis, corresponds to the location of bar corotation.

\citet{Buta2017} used multi-band images of ringed, barred galaxies to identify these `dark-gaps', and compared the results with an N-body simulation from \citet{Schwarz1984}. 
`Dark-gaps' were attributed to a deficiency of material towards $L_{4,5}$, the often stable Lagrangian points arising from the gravitational potential of a bar that may become unstable in the presence of a strong bar \citep{GalacticDynamics}. 
\citet{Buta2017} argues that identifying gaps in the galaxy images along the bar minor-axis should correspond to the $L_{4,5}$ Lagrange points, where unstable ``banana" orbits of the stars result in a decrease in the number density of stars surrounding these points. 
\citet{Kim2016} analyzed the presence of dark gaps with other bar properties and discovered that the light deficit produced by dark gaps was larger in longer and stronger bars. 
They interpreted these findings as evidence for dark gaps produced by the redistribution of mass due to bar evolution processes. 

Here, we use state-of-the-art Milky Way-like N-body simulation results to test this proposed method directly on a more general case of a barred galaxy.
Using the simulations, we construct synthetic, face-on maps of the stellar surface density and identify bar resonance locations with these gaps. 
We show the location of `dark-gaps' in this nonringed galaxy does not correspond to the $L_{4,5}$ Lagrangian points but instead occurs close to the location of the 4:1 ultra-harmonic resonance.

In this work, an alternative to the \citep{Buta2017} corotation interpretation of dark gaps is suggested based on results of
high-resolution N-body simulations \citep[GALAKOS;][]{D'Onghia2020}. 
This paper is structured as follows. Section \ref{sec:obs} briefly describes the numerical simulation and observational data used to calibrate and apply this method. 
Section \ref{sec:methods} describes our new method and tests its accuracy across different physical resolutions and epochs of the simulation. 
In Section \ref{sec:res}, we apply this method to a sample of $578$ nearby galaxies to analyze the bar dynamics in this population of barred galaxies. 
Lastly, we discuss the broad implications of our results in comparison to other methods in Section \ref{sec:disc}, and we summarize in Section \ref{sec:conclusion}. 

\section{Data}\label{sec:obs}

We use results of an N-body simulation of a Milky Way-like galaxy described in \citet{D'Onghia2020} and data associated with the Sloan Digital Sky Survey (SDSS) \citep{York2000}. 
The details of these data are described below.
Throughout this work, we assume a standard cosmology of WMAP9 \citep{Hinshaw2013} implemented via astropy\footnote{\href{https://www.astropy.org/}{{https://www.astropy.org/}}} \citep{astropy}. 

\subsection{N-body Simulations}
GALAKOS is a $\sim 90$ million N-body simulation run with GADGET3. The numerical experiment follows the motion of disk stars in a Milky Way-like galaxy. The stellar disk has  structural parameters configured to correspond with the current Milky Way \citep{JBH2016}. The simulation develops self-consistent spiral patterns and a stellar bar that arises spontaneously from the stars themselves \citep{D'Onghia2013}. 
After 2.5 Gyrs of evolution, the simulated galaxy develops prominent spiral structure and a bar with 4.5 kpc length, but does not develop a ring.

The GADGET3 cosmological code allows for fast computation of the long-range gravitational field on a particle mesh, with short-range forces calculated on a tree-based hierarchical multipole expansion. 
A spline kernel with scale length, $h_s$, softens pairwise particle interactions so that interactions beyond that scale are strictly Newtonian. 
This is similar to Plummer softening with a scale length $\epsilon = \nicefrac{h_s}{2.8}$. 
In this simulation, $h_s = \unit[40]{pc}$, $\unit[28]{pc}$, and $\unit[80]{pc}$ for the dark matter halo, stellar disk, and bulge, respectively. Full details of this simulation and the code used can be found in \citep[][and their Appendix]{D'Onghia2020}.
Our interpretation of the bar and spiral structure properties is based on the density wave theory and is supported by our numerical simulations \citep[see][]{D'Onghia2013,D'Onghia2020}.

\subsection{SDSS-IV MaNGA}
The SDSS-\Romannum{4} \citep{Blanton2017} Mapping Nearby Galaxies at Apache Point Observatory (MaNGA) survey observes nearly $10,000$ galaxies with integral field spectroscopy \citep{Bundy2015}. MaNGA observations employ the BOSS spectrograph on the 2.5 meter telescope at Apache Point Observatory \citep{Gunn2006}, achieving a spectral resolution of $R \sim 2000$ for $\unit[3600]{\text{\AA}} < \lambda < \unit[10300]{\text{\AA}}$ and a typical (S/N) level of 36 (20) at a fiducial fiber magnitude of 21 (22) in the red (blue) spectrograph arm \citep{Bundy2015}. Optical fibers subtend $\unit[2]{"}$ on the sky \citep{Smee2013}, and are bundled into integral field units (IFUs) with sizes in the range $\unit[12]{"} - \unit[32]{"}$ in diameter ($19 - 127$ fibers: \citealt{Drory2015}).
Sky subtraction and flux calibration are accomplished using simultaneous observations of the sky and standard stars \citep{Yan2016}. The median point spread function of the resulting data cubes is $\unit[2.5]{"}$ and roughly corresponds to kiloparsec physical scales at the targeted redshift range ($0.01< z < 0.15$). Observations are dithered and mapped onto $\unit[0.5]{"}$ spectroscopic pixels (or spaxels).

The MaNGA sample is selected to have a flat distribution of i-band absolute magnitude and uniform radial coverage. This parent sample is composed of three main components, a primary sample where $80\%$ of galaxies are covered out to $1.5\ R_e$, a secondary sample where $80\%$ of galaxies are covered out to $2.5\ R_e$, and a 
color-enhanced supplement to improve coverage of poorly sampled regions of the $NUV-i$ vs. $M_i$ color-magnitude plane \citep{Wake2017}. 
All MaNGA data in this work are reduced using \texttt{v2\_5\_3} of the MaNGA Data Reduction Pipeline \citep[DRP;][]{Law2016}, and employ the Data Analysis Pipeline \citep[DAP;][]{Westfall2019,Belfiore2019} from the internal eighth MaNGA product launch (MPL-8), containing $6507$ galaxies.

\subsubsection{Barred Galaxy Sample - Galaxy Zoo:3D}

Galaxy Zoo:3D \citep{Masters2021} is a citizen science project in which masks were drawn onto galaxy images to outline morphological features such as spiral arms, bars, galaxy centers, and foreground stars. After at least 15 citizen scientists have classified a galaxy, the combined masks allow for quick identification of the stellar bar and estimates of its length and position angle. 
These masks effectively separate MaNGA spaxels dominated by bar light from the rest of the galaxy. Spaxels are considered to be in the bar if so flagged by 40\% of participants. This threshold is more strict than those used in previous work \citep{Fraser-McKelvie2019,Krishnarao2020b}, but provide more accurate bar length measurements \citep[see][their Appendix A]{Krishnarao2020b}. 
Details of this citizen science project can be found on their website\footnote{\href{https://www.zooniverse.org/projects/klmasters/galaxy-zoo-3d}{(https://www.zooniverse.org/projects/klmasters/galaxy-zoo-3d)}}. 

Our barred galaxy sample is composed of $578$ face-on, barred galaxies identified in Galaxy Zoo:3D, with minor-to-major axis ratios $\nicefrac{b}{a} > 0.6$ as estimated in the NASA-Sloan Atlas \citep[NSA;][]{Blanton2011}.

\begin{figure}[ht!]
\epsscale{1.25}
\plotone{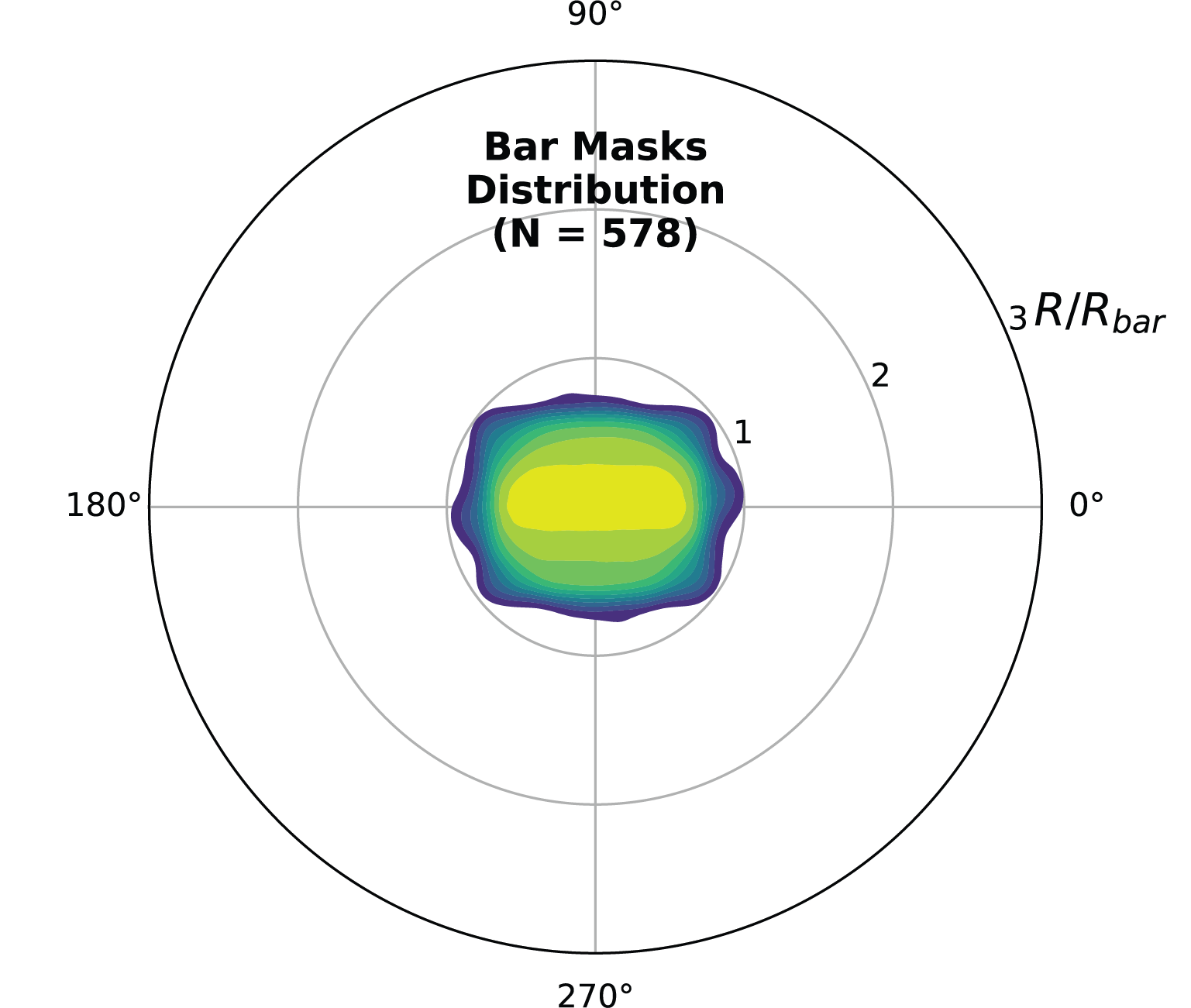}
\caption{$2$D Gaussian kernel density estimate of all bar masks in polar coordinates, normalized by bar radius.}
\label{fig:barmask_polar}
\end{figure}

The bar's orientation is determined by finding a minimum area bounding box around bar spaxels in the de-projected galaxy plane. 
Each galaxy is also re-centered based on this bounding box, and the bar length is determined using the length of this box. 
Additionally, classifications from Galaxy Zoo \citep{Lintott2011} are used to match the orientation of spiral arms in these galaxies such that the counter-clockwise direction is the direction of rotation of the spiral arm pattern, assuming all galaxies have trailing spiral arms. 
Figure \ref{fig:barmask_polar} shows a polar plot of all spaxels within the bar of the entire sample, with the bar axis orientated horizontally. 

\subsubsection{PCA Resolved Galaxy Parameters}
We use $i$-band stellar mass-to-light ratio maps for MPL-8 galaxies derived using a principal component analysis (PCA) method of stellar continuum fitting \citep{Pace2019a,Pace2019b}. 
These maps are found to be robust across a wide range of signal-to-noise ratios ($2 \le S/N \le 30$) and the full range of realistic stellar metallicities and foreground dust attenuations $(\tau \lesssim 4)$. These mass-to-light maps allow maps of the resolved stellar mass surface density to be created after de-projecting the MaNGA spaxels using the minor-to-major axis ratio from the NSA \citep{Blanton2011}, and have been previously used in \citet{Krishnarao2020b} and \citet{Schaefer2019}. 


\section{Identifying Bar Resonances}\label{sec:methods}
\citet{Buta2017} attributed `dark-gaps' along the bar minor axis to the location of bar corotation. 
These `dark-gaps' are defined as where the difference in the surface brightness along the bar major axis and the minor axis is maximal.
In what follows, we use an N-body simulation to test the validity of this proposed method in identifying the bar corotation, assuming the GALAKOS model is a realistic simulation of a barred spiral having dark gaps.

\subsection{Resonances in Simulated Barred Galaxies}
\begin{figure}[ht!]
\epsscale{1.15}
\plotone{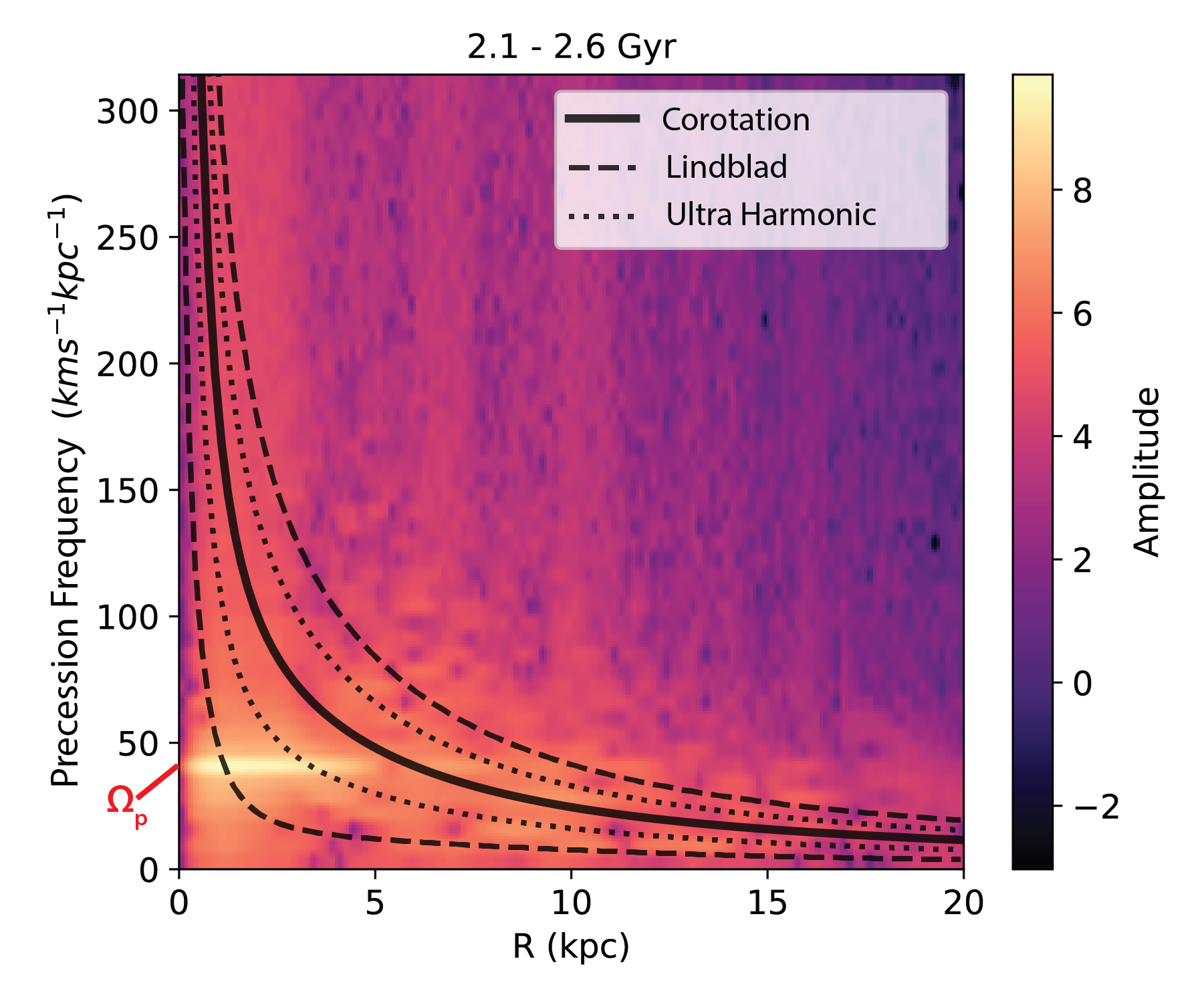}
\vspace*{-0.4cm}
\caption{$m=2$ Fourier Harmonic spectrogram from the GALAKOS simulation as a function of radius and frequencies between $2.1 < \nicefrac{t}{\textrm{Gyr}} < 2.6$, showing the bar pattern speed $\Omega_p = 39.4 \pm 0.4 \textrm{km s}^{-1} \textrm{kpc}^{-1}$. The solid line shows $\Omega = \nicefrac{V_c}{R}$, used to identify corotation and the dashed and dotted lines show similar tracks to identify the Lindblad and ultra-harmonic resonances. In this interval, $R_{\mathrm{CR}} = 6.40 \pm 0.08$ kpc and the bar patter speed, $\Omega_p$, is marked in red along the y-axis label.}
\label{fig:sim_power_spec}
\end{figure}

In the simulated galaxy, we measure the bar pattern speed, $\Omega_p$, and locations of resonances using a spectrogram as a function of radius and frequencies for the $m=2$ Fourier Harmonic in the stellar disk \citep[see][their Figure 11]{D'Onghia2020}. 
The spectrogram is constructed using snapshots sampled at $5$ Myr intervals, with the bar pattern speed corresponding to the greatest power frequency. 
Corotation corresponds to the radius at which the bar pattern speed is equal to the circular angular frequency $\Omega = \nicefrac{V_c}{R}$. 
Similarly, the Lindblad and ultra-harmonic resonances occur at the radii at which the bar pattern speed is equal to $\Omega \pm \nicefrac{\kappa}{2}$ and $\Omega \pm \nicefrac{\kappa}{4}$, respectively, where $\kappa$ is the epicyclic frequency. 
Figure \ref{fig:sim_power_spec} displays the spectrogram for m=2 applied to GALAKOS that shows the presence of a long bar with a pattern speed of $39.4 \pm 0.4 \mathrm{km s}^{-1}\mathrm{kpc}^{-1}$ as measured in the time interval between $2.1 < \nicefrac{t}{\textrm{Gyr}} < 2.6$

The radius of maximal power is identified using a quadratic fitting approach implemented using the python package, better-moments\footnote{\href{https://github.com/richteague/bettermoments}{https://github.com/richteague/bettermoments}} \citep{Teague2018} after smoothing the data using a Savitsky-Golay filter \citep{Savitzky1964} with a width of 3 radial bins. 
This allows for accurate identification, with uncertainties, of the bar pattern speed ($\Omega_p$), radius of corotation ($R_{\mathrm{CR}}$), radii of inner and outer ultra-harmonic resonance ($R_{\mathrm{iUHR}}$, $R_{\mathrm{oUHR}}$), and radii of inner and outer Lindblad resonance ($R_{\mathrm{ILR}}$, $R_{\mathrm{OLR}}$). 
We perform this analysis on $23$ time intervals of $0.25$ Gyr between $1.1<\nicefrac{t}{\textrm{Gyr}}<4.1$. 

\subsection{Simulation Stellar Surface Density}
\begin{figure}[htb!]
\epsscale{1.15}
\centering
\plotone{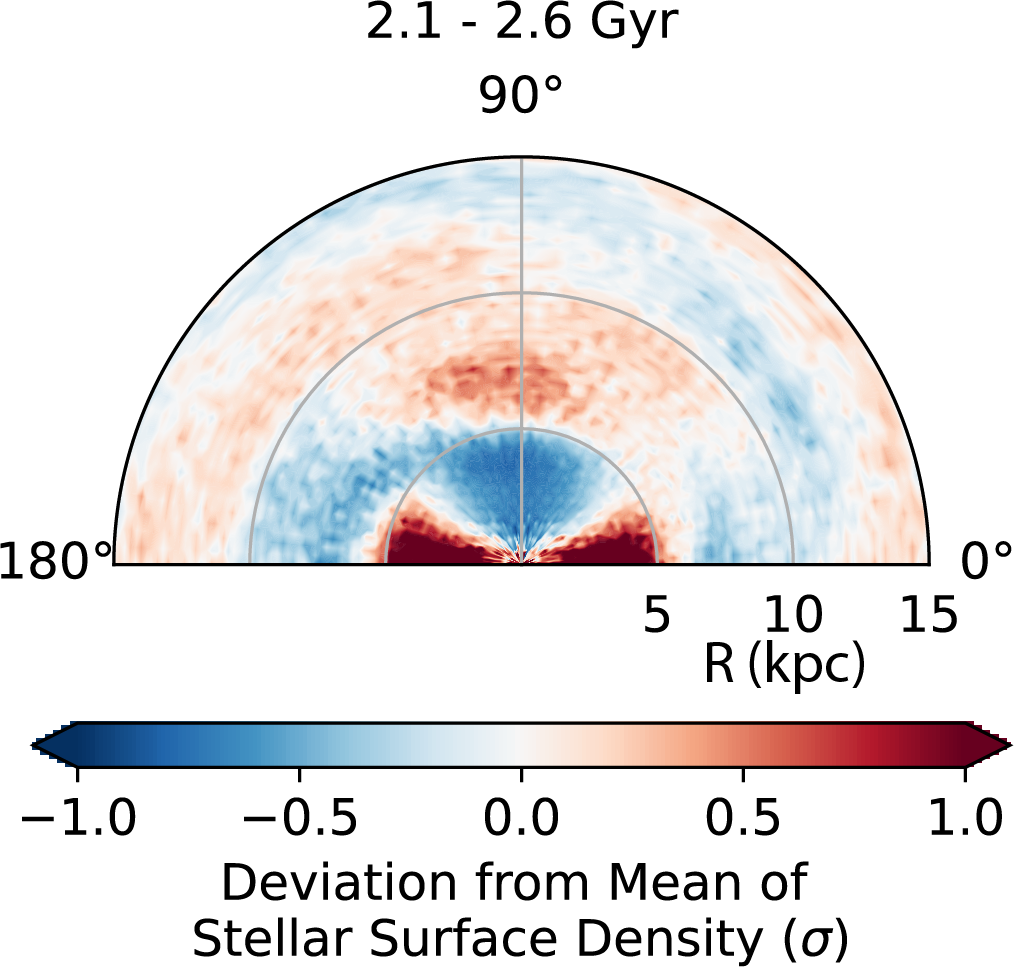}
\caption{Polar plot with same orientation as in Figure \ref{fig:barmask_polar} showing increases (red) and decreases (blue) of the stellar surface density between $2.1 < \nicefrac{t}{\textrm{Gyr}} < 2.6$ in the GALAKOS simulation. The differences from the mean values are scaled by their standard deviation, $\sigma$. A decrease in the stellar surface density is seen in the simulation along the bar minor-axis, similar to the dark-gaps used in \citep{Buta2017}.}
\label{fig:sim_polar_map}
\end{figure}

To search for the presence of dark gaps towards the $L_{4,5}$ Lagrange points and verify their correspondence with corotation, we compute the stellar surface density in the simulation, averaged across the same time intervals used to measure the pattern speeds and resonances above. 
We degrade the simulation resolution to simulate the effects of dithered MaNGA IFU observations by first taking three mock observations within dithered `beams' that sample a specified physical scale. 
Then the resulting mock observations are interpolated to finer `spaxels' that is $\nicefrac{1}{4}$ the scale of the `beams', similar to the $2"$ fibers and $0.5"$ spaxels of MaNGA.
The stellar surface density snapshots are converted to bar-centered polar coordinates, with the bar aligned along $0\degree-180\degree$. 
The bar position angle is determined using a bounding box fit around spaxels with a stellar surface density greater than $10^{-0.5}~\mathrm{pc^{-2}}$.
Next, the mock observations are binned across a radial range of $15$ kpc, with $76$ overlapping radial bins of width $0.5$ kpc. 
Similarly, the data are binned azimuthally with a width of $2\fdg5$, assuming rotational symmetry. 
We normalize these maps for each snapshot by first considering the mean value across all azimuth in a radial bin and finding the deviation from this mean value for each azimuthal bin relative to that bin's standard deviation. 
The snapshot maps are then averaged together using $10,000$ bootstrap samples, with a resulting azimuthal variation map displayed in Figure \ref{fig:sim_polar_map} for the same time interval as in Figure \ref{fig:sim_power_spec}. 

\begin{figure*}[htb!]
\epsscale{1.15}
\centering
\hspace*{-.25cm}  
\plotone{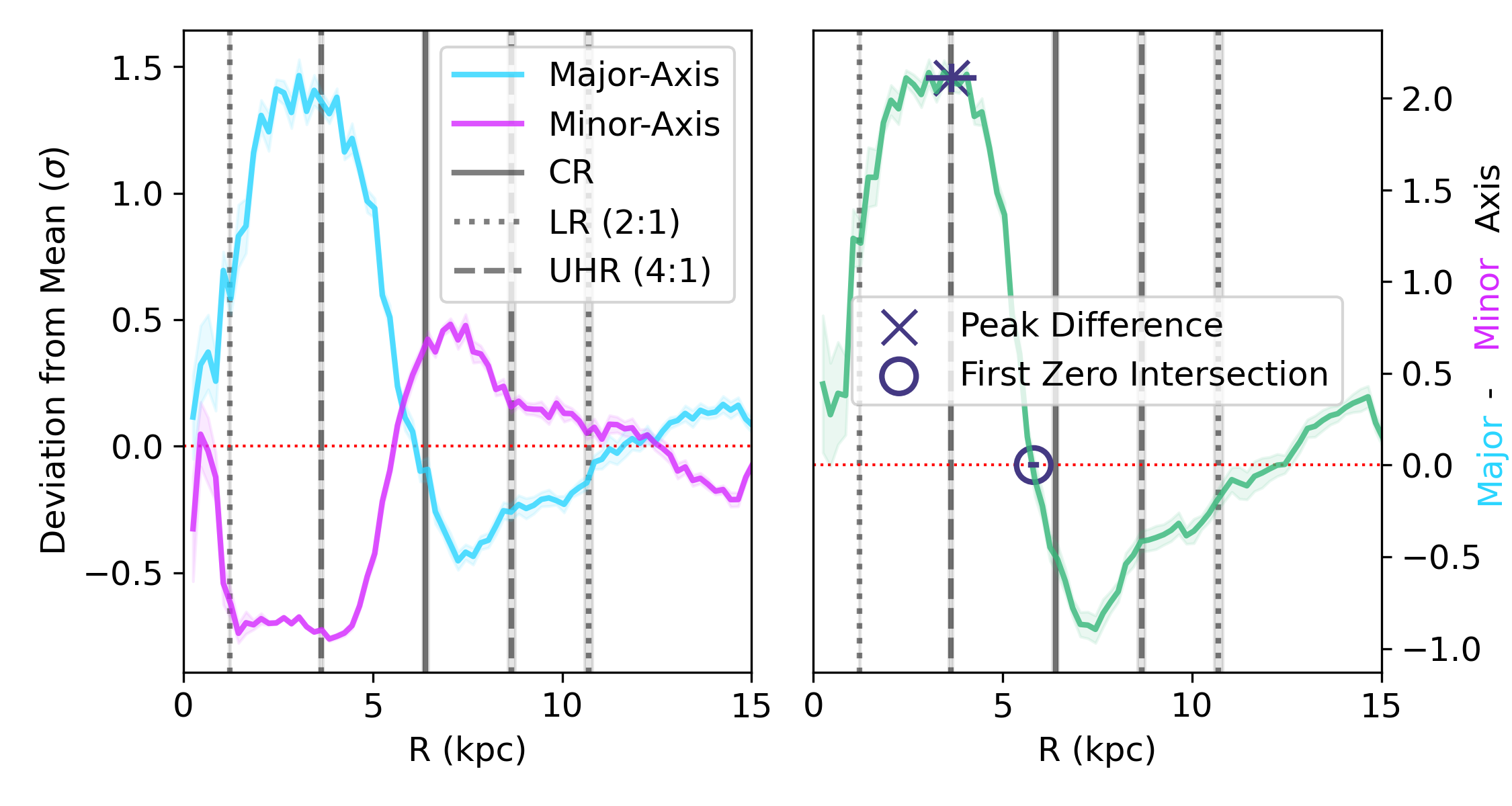}
\vspace*{-0.5cm}
\caption{(left:) Stellar surface density deviation as a function of radius between $2.1 < \nicefrac{t}{\textrm{Gyr}} < 2.6$ and within $30\degree$ of the bar major (blue) and minor (pink) axis from the GALAKOS simulation. (right:) The difference between the major and minor axis surface density deviations as a function of radius. The radii with a maximal difference and a null difference are marked with an X and O, respectively. In both panels, the solid, dotted, and dashed black vertical lines mark the locations of corotation, the Lindblad resonances, and the ultra-harmonic resonances, respectively.}
\label{fig:sim_dev_plot}
\end{figure*}

Figure \ref{fig:sim_dev_plot} shows the relative increase and decrease of the stellar surface density within $30\degree$ of the bar major and minor axes, respectively, again for the same time interval.
The difference between the major and minor axes' radial stellar mass profiles and the peak difference location is shown on this Figure's right panel. 
\citet{Buta2017} attributed this effect to the $L_{4,5}$ and $L_{1,2}$ Lagrange points, and thus approximately at the radius of corotation. 
However, this radius does not correspond to corotation in the simulation, but instead is the inner ultra-harmonic bar resonance location.
In this example, corotation is located at a larger galactocentric radius, near the inner-most radius where the differences between the major and minor axes are zero, which we call $R_\mathrm{Cross,CR}$. 
Note that this metric is not always accurate and tends to underestimate the correct corotation radius slightly. 

These synthetic observations of a simulated barred disk galaxy provide an interesting alternative to the dark gap/corotation interpretation described by \citep{Buta2017}. The residual gap in the GALAKOS model is found to coincide with the model's inner ultra-harmonic, not its corotation.
\citet{Athanassoula1982} showed that the ratio of the radii of corotation and the ultra-harmonic resonance can be written as
\begin{equation}
    \left(\frac{R_{\mathrm{CR}}}{R_{\mathrm{UHR}}}\right)^\delta = \frac{1}{1 - \frac{1}{2} \Delta}
    \label{master_eq}
\end{equation}
where $\Delta = \left( 1 - \frac{1}{2}\delta \right)^{\nicefrac{1}{2}}$ and $\delta$ describes the galaxy rotation curve where $V \sim r^{-\delta + 1}$.
$\delta = 1$ corresponds to a flat rotation curve, and \citet{Athanassoula1982} notes that $0.7 \leq \delta \leq 1.0$ is representative of most barred galaxies. 
We therefore can estimate the radius of corotation based on the ultra-harmonic resonance, which we call $R_\mathrm{Ratio,CR}$, as
\begin{equation}
    \nicefrac{R_{\mathrm{Ratio,CR}}}{R_{\mathrm{UHR}}} = 1.8\pm0.3
    \label{eq1}.
\end{equation}
Alternatively, corotation may also correspond with the inner-most radius where the major to minor axis differences are zero ($R_\mathrm{Cross,CR}$; labeled as ``First Zero Intersection'' in Figure \ref{fig:sim_dev_plot}), though we later show that this method is not always accurate.

\begin{figure*}[htb!]
\epsscale{1.15}
\centering
\hspace*{-.25cm}  
\plotone{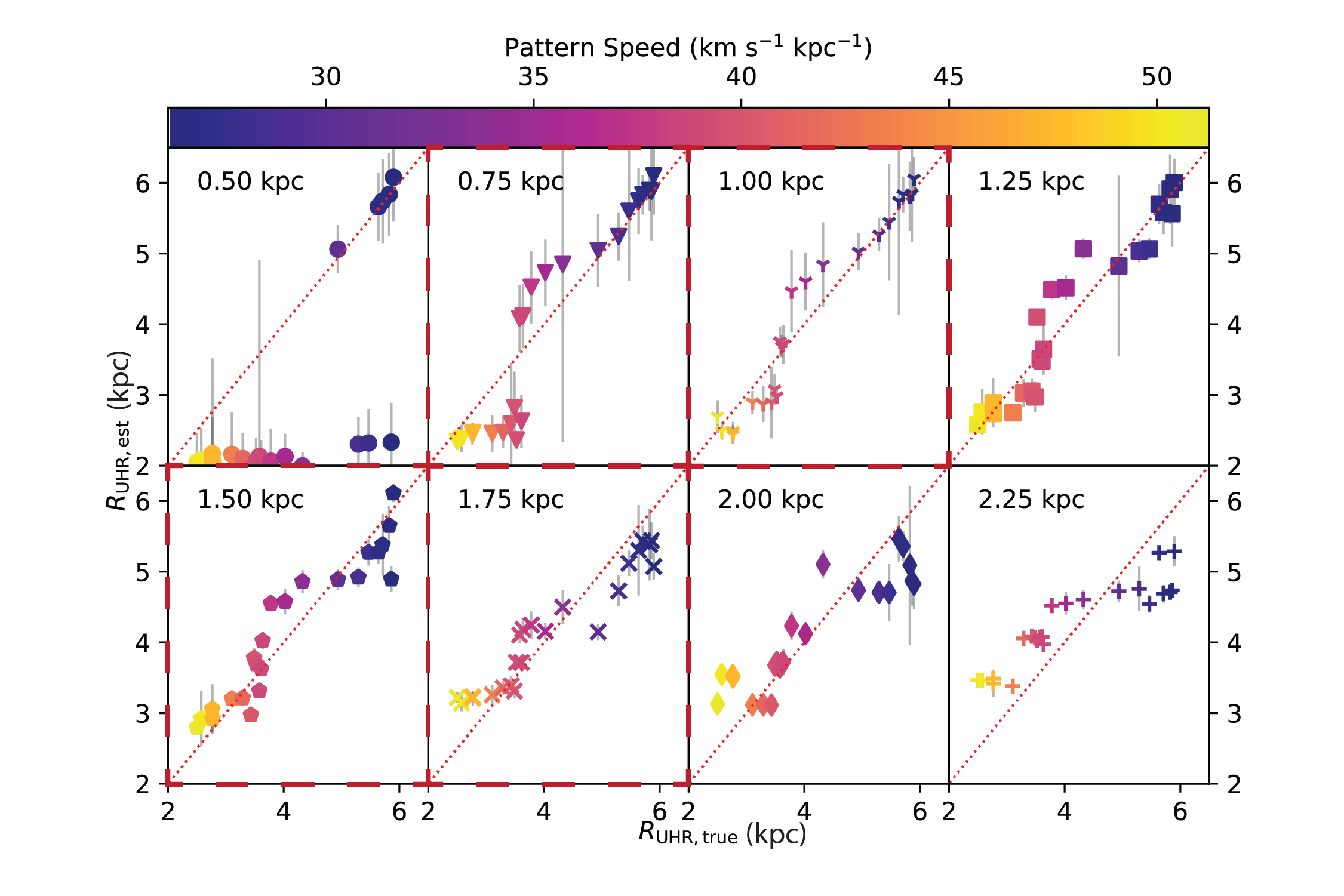}
\vspace*{-0.5cm}
\caption{The estimated vs. true UHR radius across $23$ time intervals of $09.25$ Gyr in the range of $1.1<\nicefrac{t}{\textrm{Gyr}}<4.1$ in the GALAKOS simulation, with points color-coded by the bar pattern speed at each snapshot. Each panel corresponds to a different physical spaxel resolution (labeled in the top-left of each panel) of the stellar density images used in the measurements, with red dashed outlines highlighted resolution scales corresponding to MaNGA galaxies. The red dotted line shows a one-to-one relation. }
\label{fig:UHR_estVtrue}
\end{figure*}

\begin{figure*}[htb!]
\epsscale{1.15}
\centering
\hspace*{-.25cm}  
\plotone{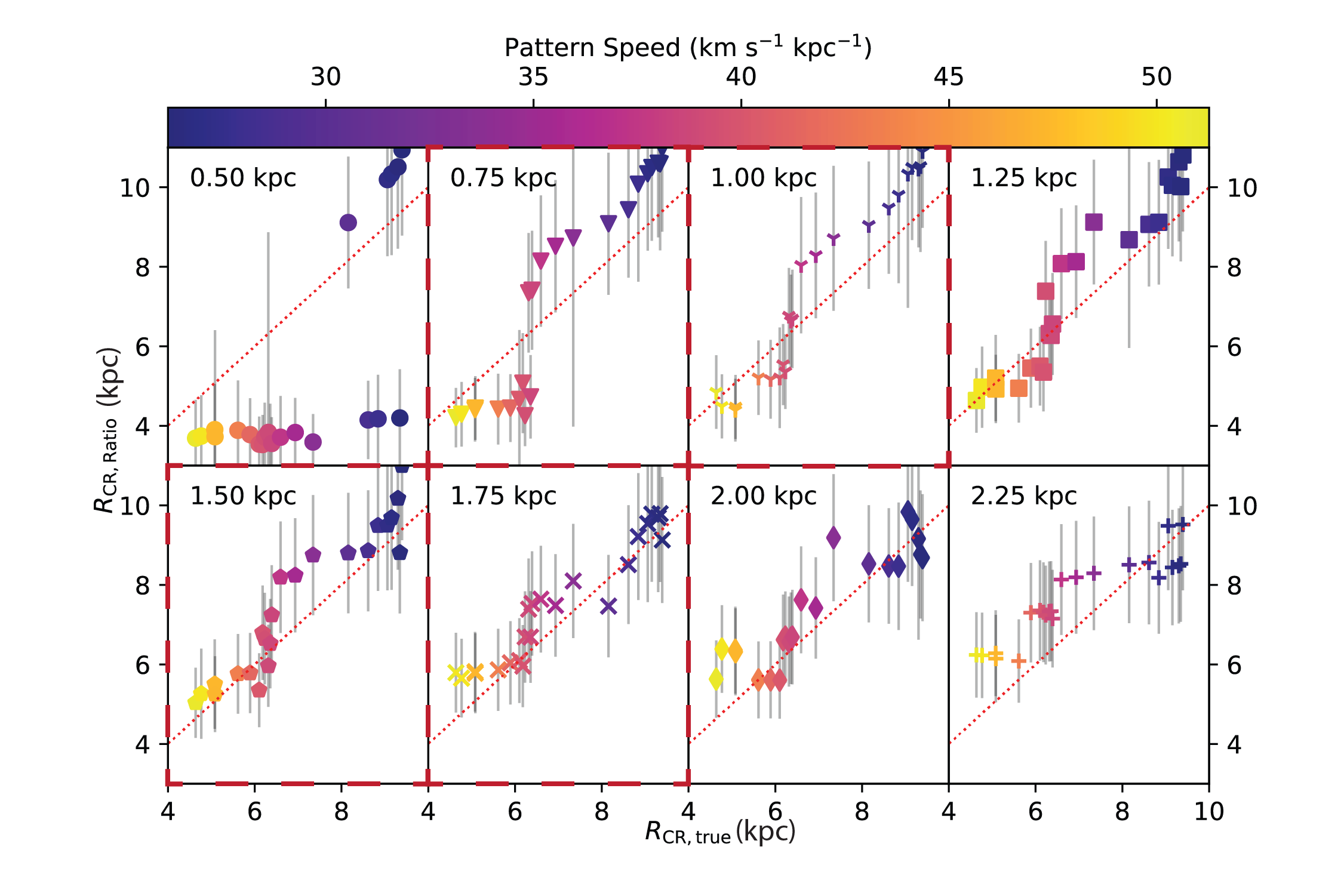}
\vspace*{-0.5cm}
\caption{Our estimated vs. true corotation radius assuming a range of rotation curve shapes across $23$ time intervals of $09.25$ Gyr in the range of $1.1<\nicefrac{t}{\textrm{Gyr}}<4.1$ in the GALAKOS simulation, with points color-coded by the bar pattern speed at each snapshot. Each panel corresponds to a different physical resolution (labeled in the top-left of each panel) of the stellar density images used in the measurements, with red dashed outlines highlighted resolution scales corresponding to MaNGA galaxies. The red dotted line shows a one-to-one relation. }
\label{fig:CR_Ratio_estVtrue}
\end{figure*}

We then test the ability of our dark-gap-based method to identify the UHR throughout the dynamical evolution of our simulated galaxy. Figure \ref{fig:UHR_estVtrue} shows the estimated and true radii of the UHR across $23$ time intervals of $0.25$ Gyr that overlap in the range of $1.1<\nicefrac{t}{\textrm{Gyr}}<4.1$ and for different physical resolutions of the stellar density images. 
Over this time interval, the bar evolves and decreases its pattern speed through angular momentum transport.
Our method tends to work best when considered at the approximate physical resolution of MaNGA spaxels, where at higher resolution, finer density variations add significant noise and local extrema. 
Figure \ref{fig:CR_Ratio_estVtrue} shows the same methodology but when applied for corotation as estimated using Equation \ref{eq1}. The uncertainties associated with $R_\mathrm{Ratio,CR}$ are large here because we assume a large range in the $\delta$ parameter that describes the rotation curve shape.

\begin{figure*}[htb!]
\epsscale{1.15}
\centering
\hspace*{-.25cm}  
\plotone{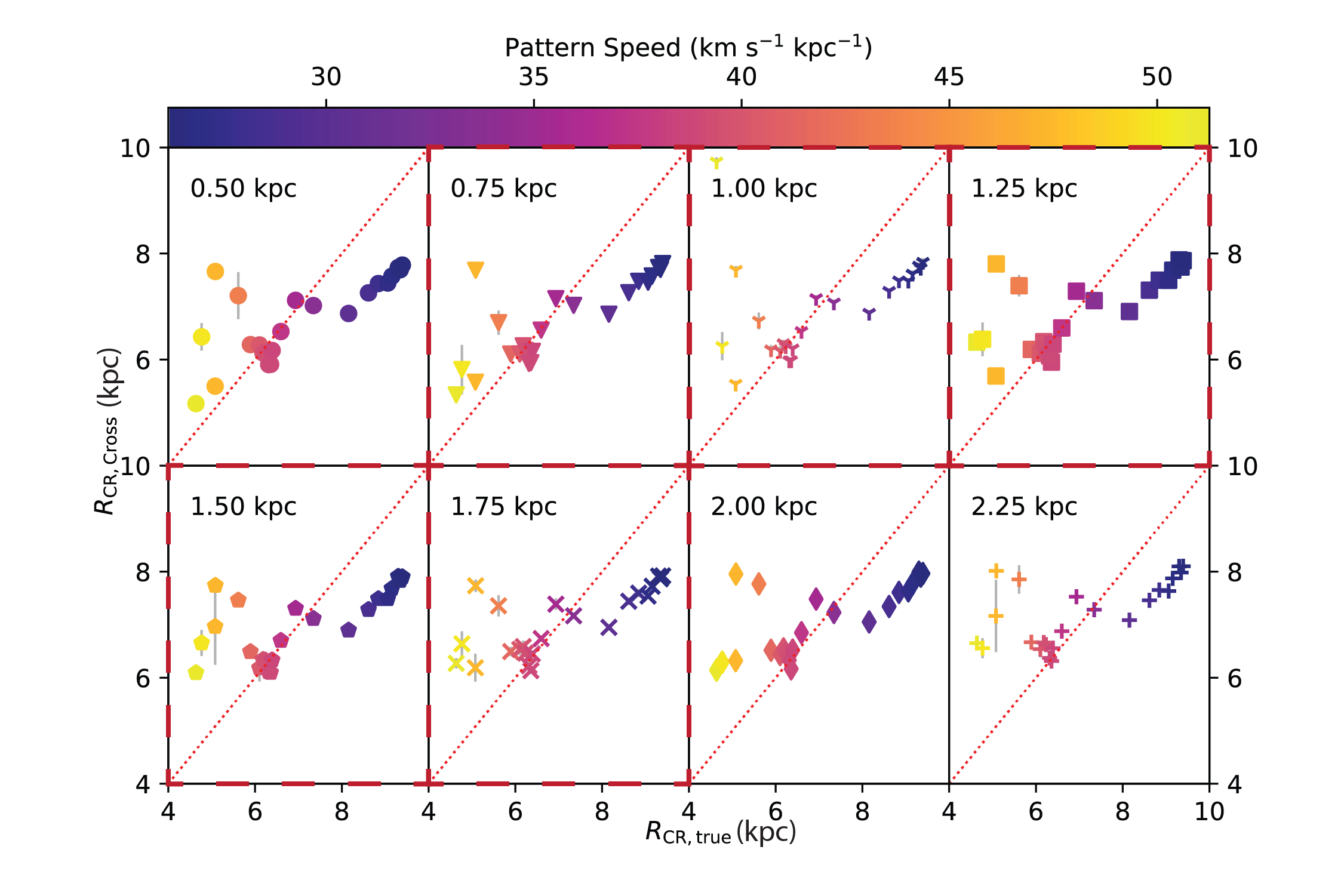}
\vspace*{-0.5cm}
\caption{The estimated vs. true corotation radius using the inner-most radius where the major to minor axis differences are zero across $23$ time intervals of $09.25$ Gyr in the range of $1.1<\nicefrac{t}{\textrm{Gyr}}<4.1$ in the GALAKOS simulation, with points color-coded by the bar pattern speed at each snapshot. Each panel corresponds to a different physical resolution (labeled in the  top-left of each panel) of the stellar density images used in the measurements, with red dashed outlines highlighted resolution scales corresponding to MaNGA galaxies. The red dotted line shows a one-to-one relation. }
\label{fig:CR_Cross_estVtrue}
\end{figure*}

Figure \ref{fig:CR_Cross_estVtrue} shows the same again, but for an estimate of corotation using $R_\mathrm{Cross,CR}$, the inner-most radius where the major to minor axis differences are zero. This method does not adequately predict the corotation radius's valid location across different evolutionary phases of the galaxy. Instead, the location of this cross point is strongly affected by the bar pattern speed. 

\begin{figure*}[htb!]
\epsscale{1.15}
\centering
\plotone{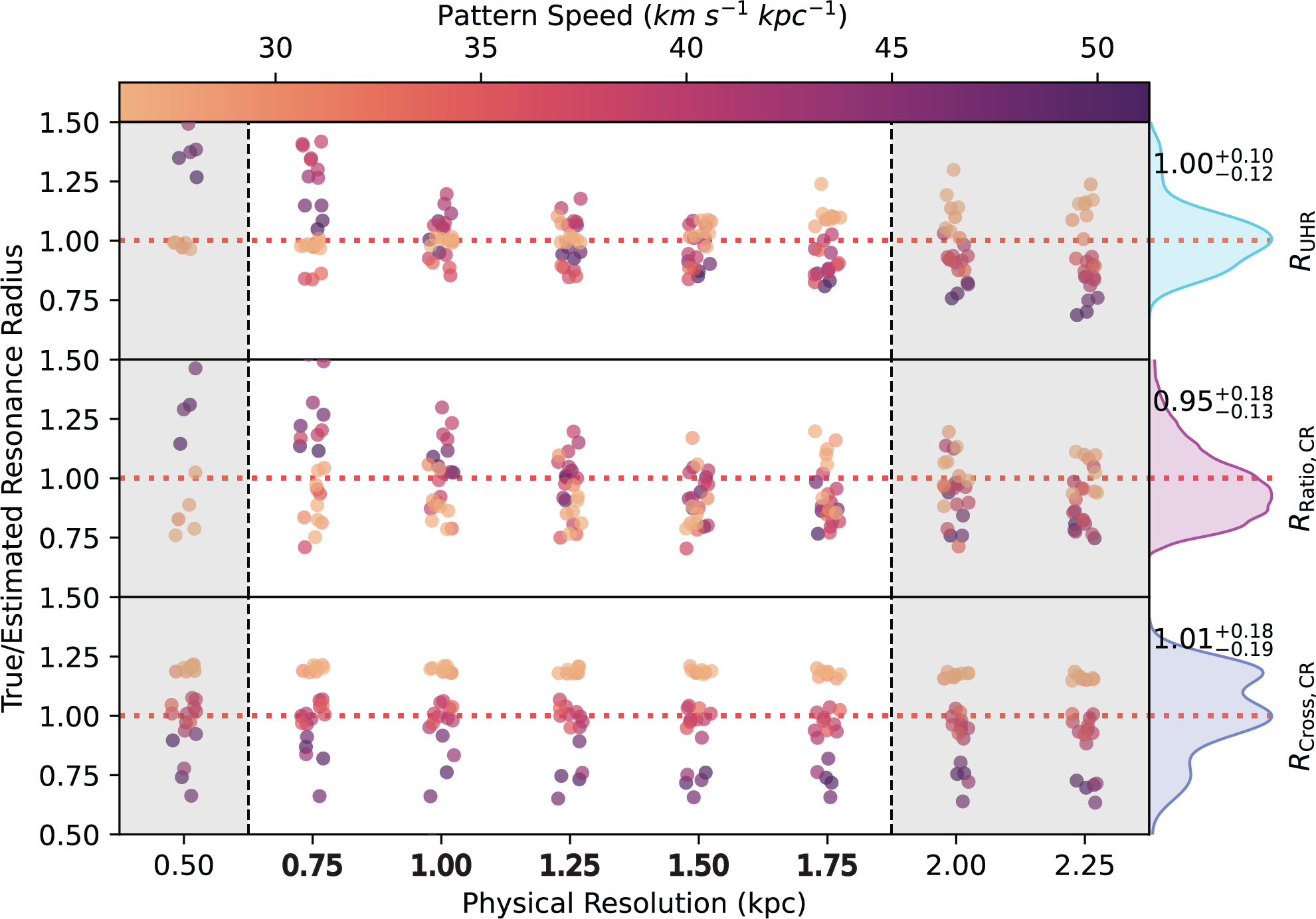}
\caption{left: True/estimated radii of the ultra-harmonic resonance ($R_\mathrm{UHR}$; upper panel) and corotation as measured using a ratio ($R_\mathrm{Ratio,CR}$; middle panel) and using the first-zero crossing point ($R_\mathrm{Cross,CR}$; lower panel) at different physical resolution scales in 23 intervals of $0.25$ Gyr between $1.1<\nicefrac{t}{\textrm{Gyr}}<4.1$. The pattern speed of the bar colors the points in the corresponding simulation snapshot. The approximate physical scale of the $2"$ MaNGA fibers span the resolution scales in bold. A precise and accurate estimate would be narrowly distributed and centered on one (dashed-red lines). Right: Histograms of the true/estimated radii for the resolutions comparable to MaNGA galaxies. The median ratios and uncertainties estimated using the $16th$ and $84th$ percentiles of the distributions are displayed for each metric. These will be used as a correction factor.}
\label{fig:sim_UHR_CR}
\end{figure*}

Figure \ref{fig:sim_UHR_CR} shows all three of these metrics in terms of a fractional error (True/Estimated) across different time intervals and physical resolution scales. Within the range of resolutions corresponding with MaNGA galaxies, we can describe the accuracy and precision for identifying the UHR as $\frac{R_\mathrm{UHR,True}}{R_\mathrm{UHR,Est}} = 1.00^{+0.10}_{-0.12}$. Similarly by assuming a rotation curve shape we find $\frac{R_\mathrm{CR,True}}{R_\mathrm{Ratio,CR}} = 0.95^{+0.18}_{-0.13}$. Although $R_\mathrm{Cross,CR}$ is not an ideal estimator, it returns an averaged accuracy of $\frac{R_\mathrm{CR,True}}{R_\mathrm{Cross,CR}} = 1.01^{+0.18}_{-0.19}$. We adopt these simulation derived-ratios as a correction factor, with uncertainties, to adjust the measured locations and improve the accuracy of the predicted resonance locations. 
With this method validated on simulation data and with an understanding of the corrections needed to be made and uncertainties involved, we can apply it to a sample of observed galaxies, both ringed and non-ringed, to better understand its validity. 

\section{Observational Results}\label{sec:res}

The $578$ barred galaxies with bar masks and resolved galaxy parameters provide a unique statistical view into the impact of a bar on galaxies' distribution of matter. 
To consider our sample as a whole and understand the significance of rings in our barred galaxies, we split our sample into two sub-samples of $211$ ringed and $367$ nonringed galaxies using a 50\% unbiased vote fraction on the Galaxy Zoo question `t08\_odd\_feature\_a19\_ring' from \citet{Hart2016}. We then project all ringed and nonringed galaxies onto a bar-oriented cylindrical coordinate system, with the radius in units of the bar radius ($R_{\textrm{bar}}$) and azimuth angle oriented such that the bar major axis spans from $0\degree$ to $180\degree$.
We use visual classifications from Galaxy Zoo \citep{Lintott2011} to orient all spiral arms in a counter-clockwise spiral pattern. In this bar reference frame, we assume a $180\degree$ rotational symmetry so that spaxels at an azimuth of $30\degree$ and $210\degree$ are equivalent. 
To ensure an equal sampling of different radii in galaxies, we restrict each galaxy to only extend out to the maximum bar-oriented radius found in all azimuthal directions. 
Additionally, we remove spaxels with a g-band signal-to-noise $< 2$ to only consider spaxels that lie on the galaxy disk. We can then search for azimuthal and radial surface density variations in the context of bar-driven resonances. 

\subsection{Surface Density Variations}

\begin{figure*}[ht!]
\epsscale{1.}
\centering
\plotone{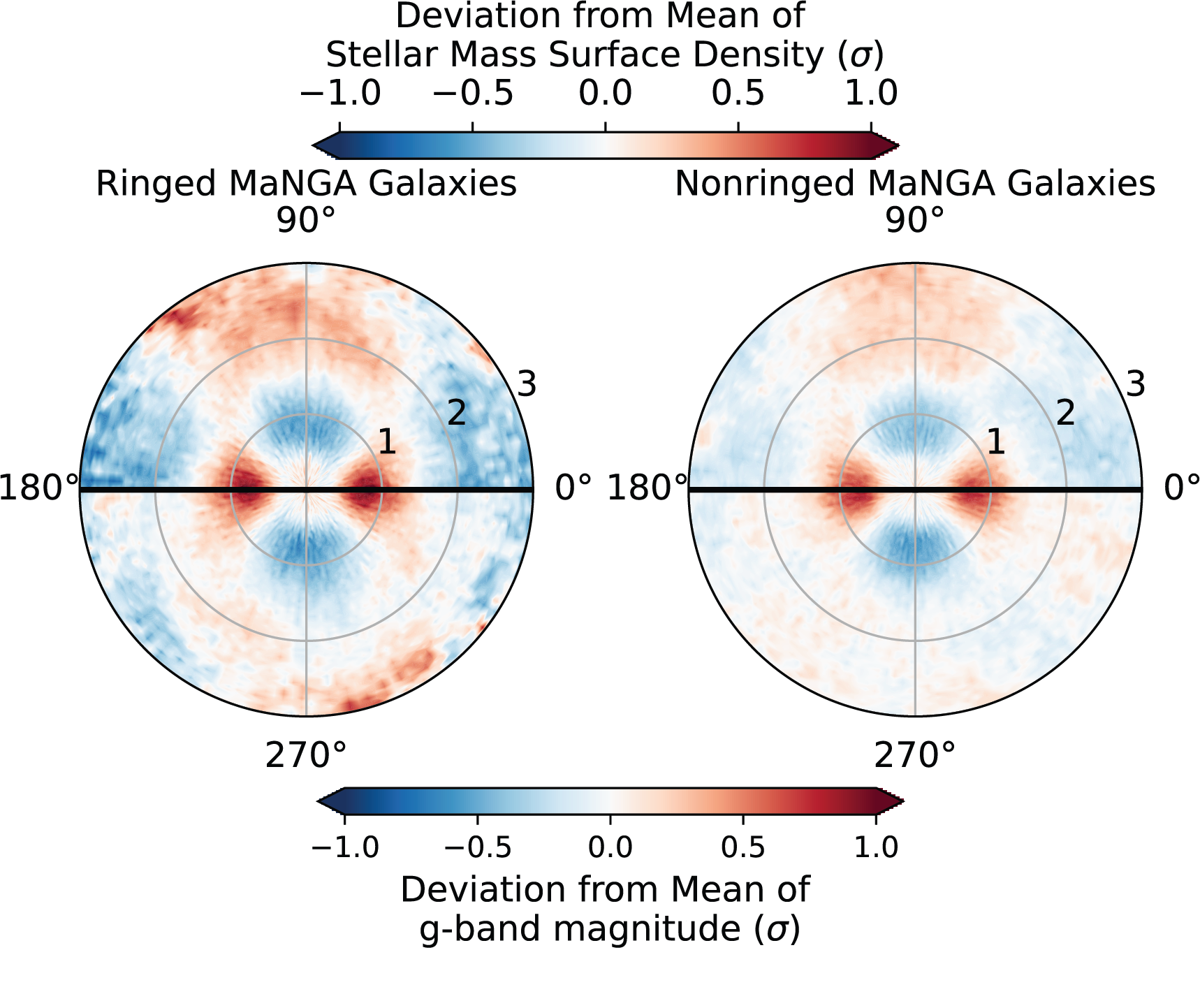}
\vspace*{-0.8cm}
\caption{Same as in Figure \ref{fig:sim_polar_map} but for the $578$ barred MaNGA galaxies using the PCA stellar mass surface density of \citet{Pace2019a,Pace2019b} (top half; $0\degree - 180\degree$) and SDSS G-band imaging (bottom half; $180\degree - 360\degree$), separated into $211$ ringed (left) and $367$ nonringed (right) galaxies. The radial axis is normalized by the bar radius.}
\label{fig:obs_polar_map}
\end{figure*}

Maps of the stellar mass surface density from \citet{Pace2019a, Pace2019b} provide a resolved measure of the distribution of stars in MaNGA galaxies. 
The maps are binned in a similar radial \& azimuthal schema as the N-body simulations: radial bins are computed relative to the bar radius, overlap with a width of $\unit[0.2]{\nicefrac{R}{R_{\textrm{bar}}}}$, and have a step size of $\unit[0.04]{\nicefrac{R}{R_{\textrm{bar}}}}$, out to a maximum radius $R = \unit[3]{R_{\textrm{bar}}}$. 
We normalize these maps for each galaxy by first considering the mean value across all azimuth in a radial bin and finding the deviation from this mean value for each azimuthal bin in terms of the standard deviation computed across all azimuth. 
We also do the same using SDSS g-band imaging to test the ability to perform this analysis using only imaging data.
This process produces maps for all $578$ galaxies, showing azimuthal variations across different radial bins, scaled by each radial bin's standard deviation. 
We combine the resulting maps using $10,000$ bootstrap re-samples to create Figure \ref{fig:obs_polar_map}. 
The stellar mass surface density and g-band magnitude show similar behavior in the simulations and Figure \ref{fig:sim_polar_map}. 

\begin{figure*}[htb!]
\epsscale{1.15}
\centering
\hspace*{-.25cm}  
\plotone{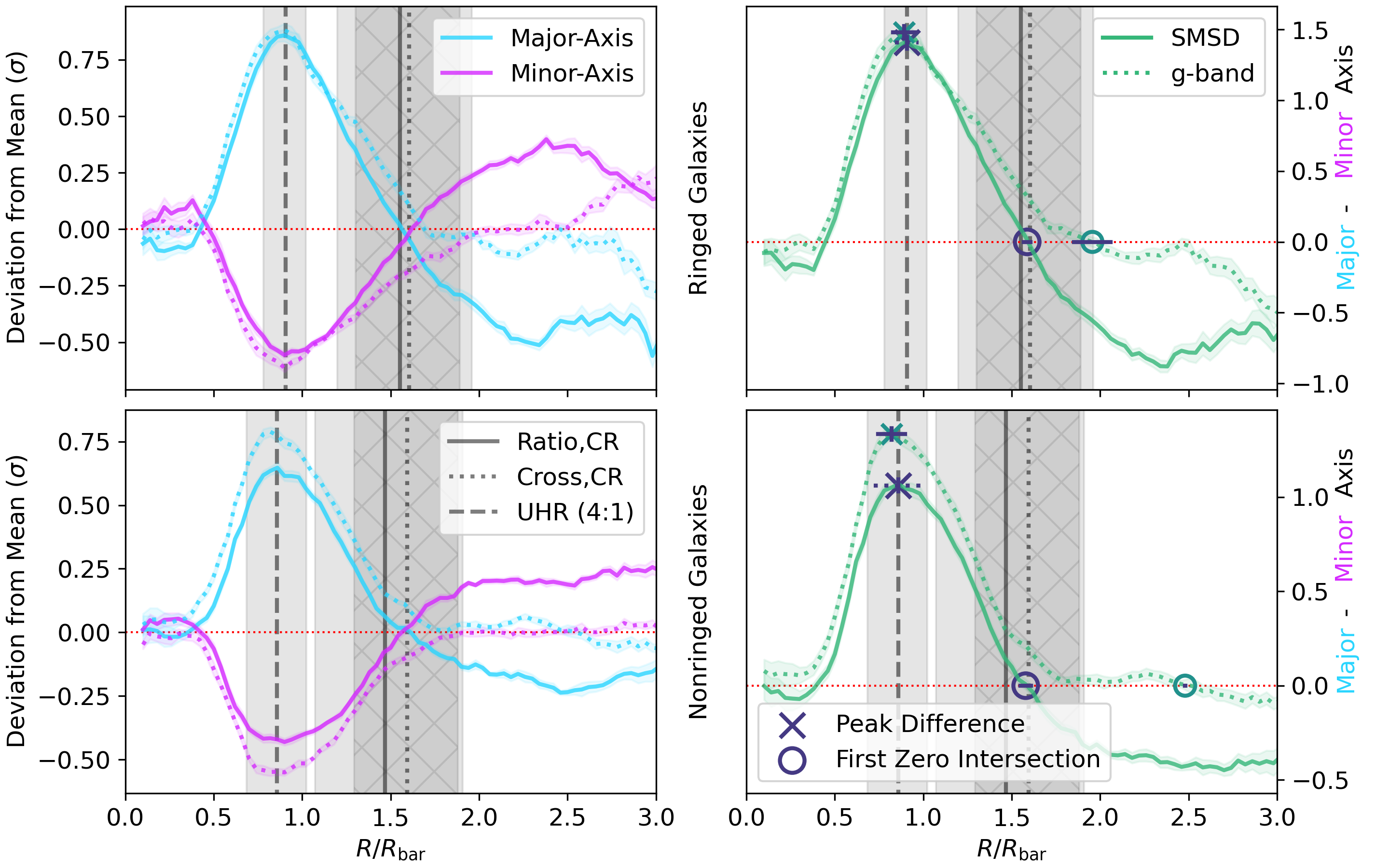}
\vspace*{-0.25cm}
\caption{Same as in Figure \ref{fig:sim_dev_plot} but for the $578$ barred MaNGA galaxies, separated into $211$ ringed (upper panel) and $367$ nonringed (lower panel) galaxies using the PCA stellar mass surface density of \citet{Pace2019a,Pace2019b} (solid-lines) and SDSS g-band imaging (dotted-lines). The inner ultra-harmonic resonance ($R_{\mathrm{UHR}}$) is approximated at the location of maximal difference between the major and minor axes and corotation is approximated as $\nicefrac{R_{\mathrm{Ratio,CR}}}{R_{\mathrm{UHR}}} = 1.8\pm0.3$ (solid shading) and as the first zero intersection ($R_\mathrm{Cross,CR}$; hatched shading). All estimates have also been adjusted to include the correction factors found in the simulations (see Figure \ref{fig:sim_UHR_CR}).}
\label{fig:obs_dev_plot}
\end{figure*}

Figure \ref{fig:obs_dev_plot} follows the same method used for Figure \ref{fig:sim_dev_plot}, but with the observed stellar mass surface density and g-band magnitudes of MaNGA galaxies. 
The peak difference is identified in the same manner, and the correction factors are applied to adjust for the expected offsets based on the simulation (see Figure \ref{fig:sim_UHR_CR}).
We find the average location of the inner ultra-harmonic resonance to be $R_{\mathrm{UHR}} = 0.89_{-0.13}^{+0.11} R_{\mathrm{bar}}$ with estimated corotation radii with both methods of $R_\mathrm{Cross,CR} = 1.60^{+0.29}_{-0.30} R_{\mathrm{bar}}$ and $R_\mathrm{Ratio,CR} = 1.52^{+0.40}_{-0.35} R_{\mathrm{bar}}$. 
Bars with $\nicefrac{R_\mathrm{CR}}{R_{bar}} \geq 1.4$ are considered ``slow", while bars with $\nicefrac{R_\mathrm{CR}}{R_{bar}} < 1.4$ are considered ``fast" \citep{Debattista2000}. 
Bars with $\nicefrac{R_\mathrm{CR}}{R_{bar}} < 1$ are often referred to as ``ultrafast", but it is unclear whether these bars are a real feature or a result of imperfect estimates of either $R_\mathrm{CR}$ or $R_\mathrm{bar}$ \citep[e.g.][]{Buta2017}.
Our results suggest that slow bars are the most likely scenario for this sample of galaxies as an ensemble, since $R_\mathrm{CR}/R_\mathrm{bar} > 1.4$.
The sample of galaxies considered in \citet{Buta2017} are also primarily slow, with $R_\mathrm{CR}/R_\mathrm{bar} \geq 1.58$.
However, the relatively large uncertainties in this population-level diagnostic do not categorically exclude the fast rotator alternative. 

\begin{figure*}[htb!]
\epsscale{1.15}
\centering
\hspace*{-.25cm}  
\plottwo{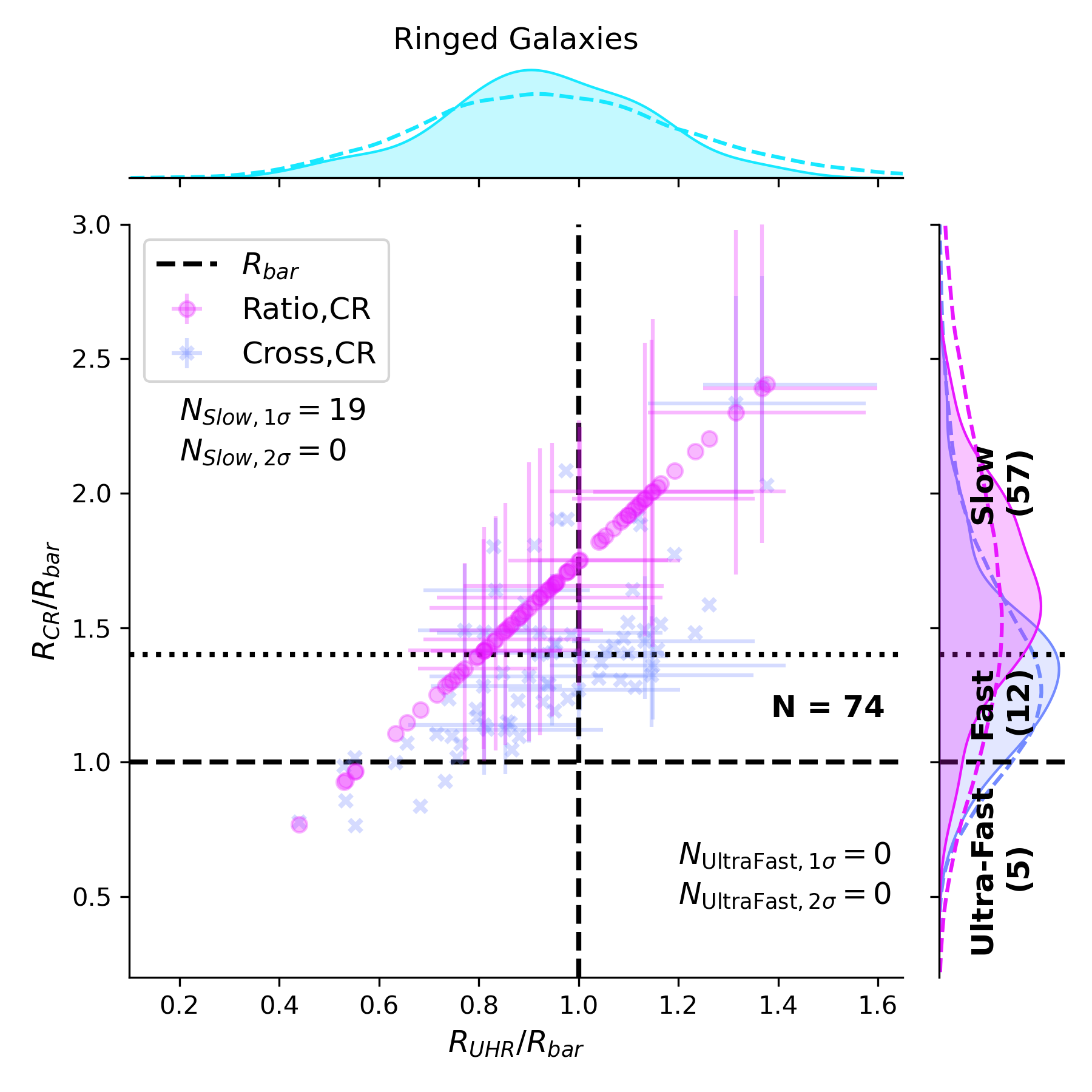}{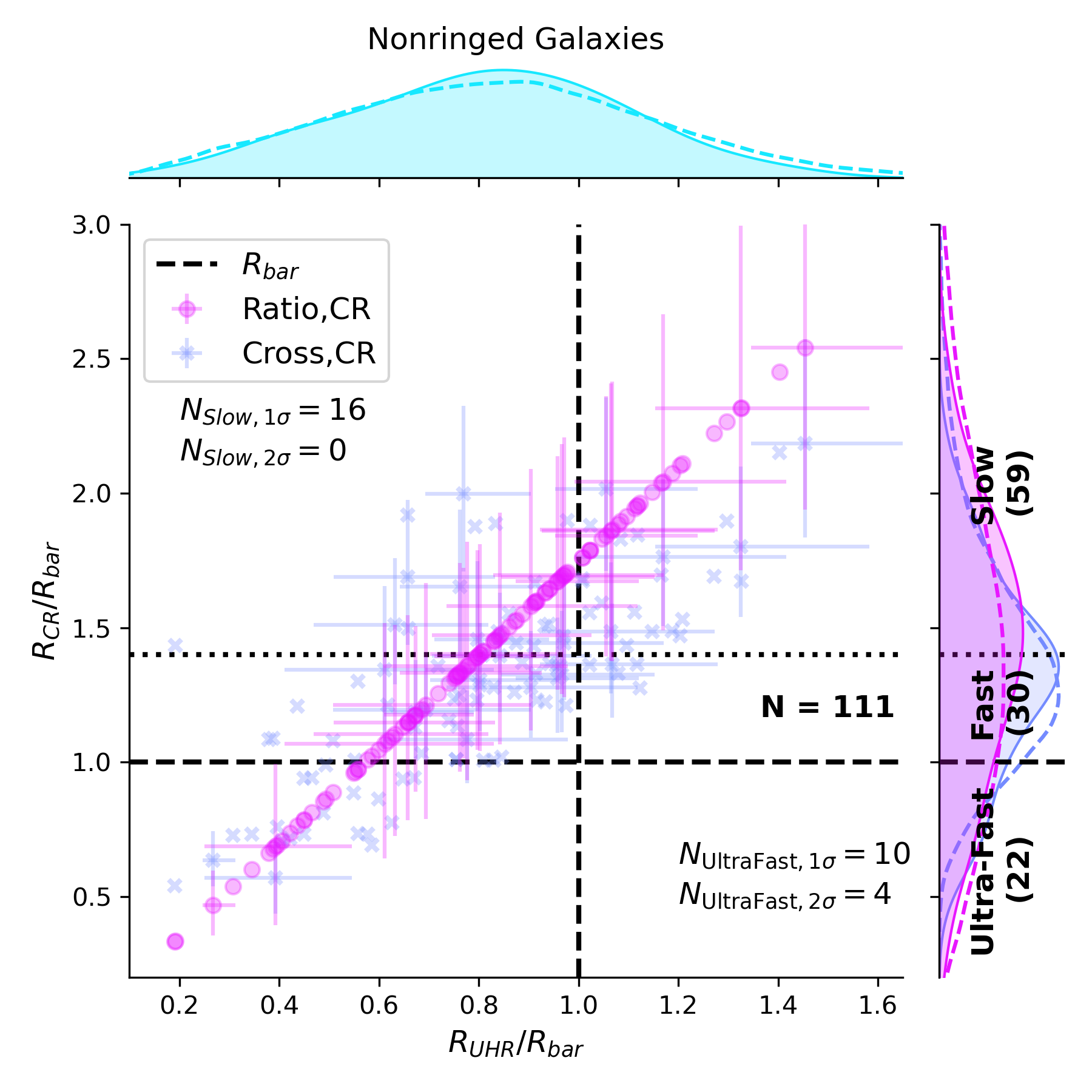}
\vspace*{-0.2cm}
\caption{Estimated radius of the inner ultra-harmonic resonance ($R_{\mathrm{UHR}}$) and radius of corotation ($R_{\mathrm{CR}}$) for individual ringed (left) and nonringed (right) MaNGA galaxies. Data are limited to galaxies where measurement errors are under $0.2 R_{bar}$ for identifying the peak difference. Error bars are shown for one-fifth of the data points. The dotted horizontal lines mark the boundaries between ultra-fast, fast, and slow bars. Marginalized Gaussian kernel density estimates are shown along both axes, with the dashed lines including the associated errors. A total of $74$ ringed and $111$ nonringed galaxies are plotted, with $5$ ultra-fast, $12$ fast and $57$ slow bars in the ringed sample and $22$ ultra-fast, $30$ fast and $59$ slow bars in the nonringed sample as diagnosed using $R_{\mathrm{Cross,CR}}$. When considering the errors, the number of galaxies with ultra-fast and slow bars drops significantly. }
\label{fig:obs_all_CR}
\end{figure*}

\begin{figure*}[htb!]
\epsscale{1.15}
\centering
\hspace*{-.25cm}  
\plotone{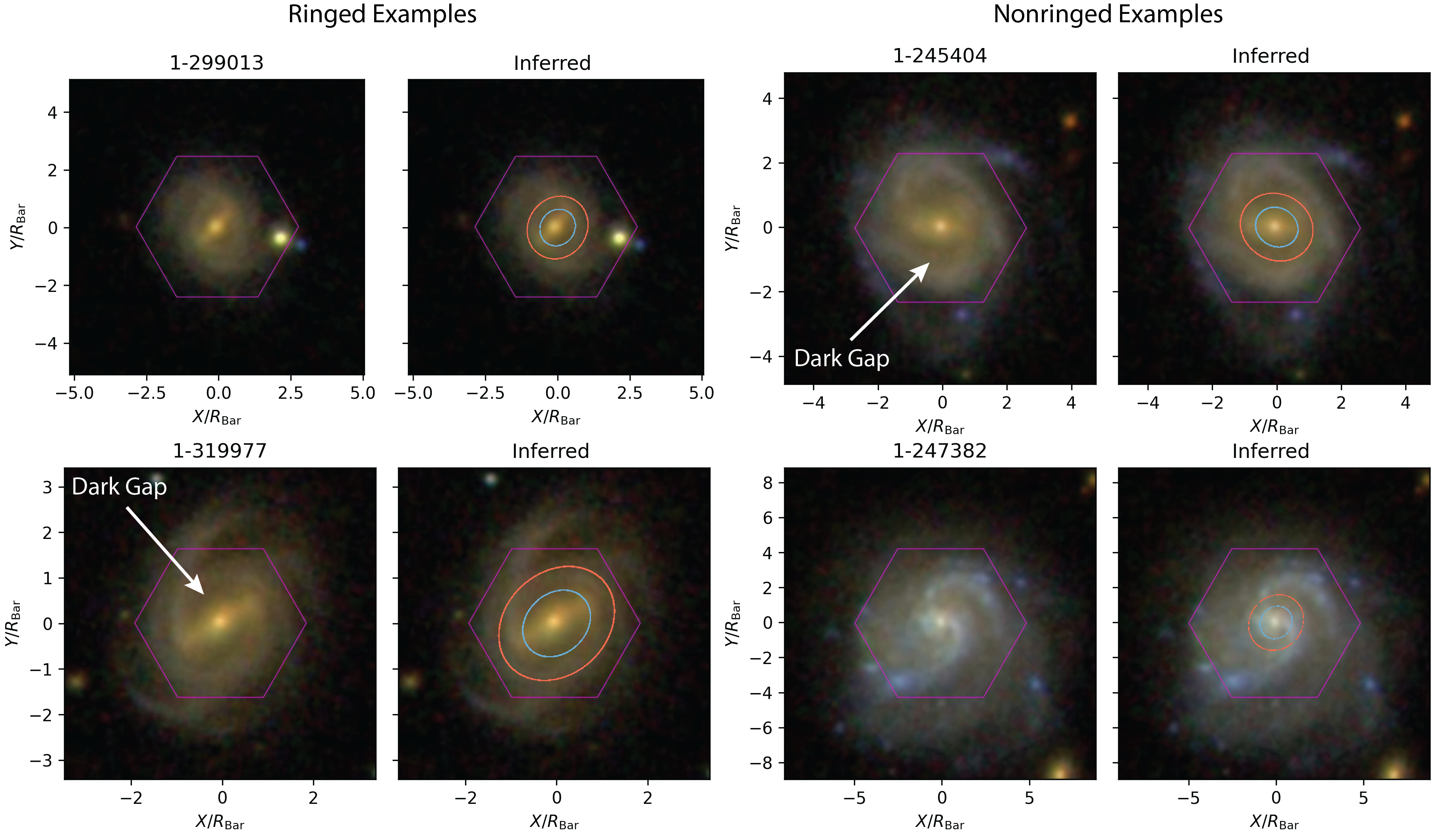}
\vspace*{-0.2cm}
\caption{Example SDSS gri images with the inferred location of the inner ultra-harmonic resonance and corotation shown on the right-hand image pair in blue and red, respectively, for ringed (left) and nonringed (right) galaxies. The image axes are scaled based on the measured bar radius. The MANGA-ID of each galaxy is displayed above the left-hand image pair. The purple hexagons mark the footprint of the MaNGA survey. These images are exactly what are used in the citizen science classifications of \citet{Masters2021}.}
\label{fig:sdss-images}
\end{figure*}

We also perform the same analysis on the $578$ individual galaxies considered here, with $185$ galaxies returning reasonable estimates of $R_{\mathrm{UHR}}$ and $R_{\mathrm{CR}}$ using the PCA stellar mass surface density as shown in Figure \ref{fig:obs_all_CR}. 
Four example galaxy images with their inferred ultra-harmonic resonance location marked with shaded rings are shown in Figure \ref{fig:sdss-images}.
While many galaxies seem to be classified as slow, the large uncertainties with the corotation radius estimates leave no galaxies with a $2\sigma$ estimate of $R_\mathrm{CR}/R_\mathrm{bar}$ that is slow. Combined with the fact that the peak of the $R_\mathrm{CR}/R_\mathrm{bar}$ distribution for these $185$ galaxies is within the fast threshold, it seems likely that most bars in MaNGA galaxies in our sample are, in fact, fast rotators. This is in general agreement with an independent study using a smaller sample of MaNGA galaxies with the \citet{Tremaine1984} method, which found $\nicefrac{R_\mathrm{CR}}{R_{bar}} = 1.17^{+0.5}_{-0.41}$ \citep{Garma-Oehmichen2020}.

\section{Discussion}\label{sec:disc}
Rings are often suggested to be linked to resonances with the pattern speed of bars. 
However, measurements of the pattern speed and location of resonances are difficult to observe in large samples of galaxies, especially for face-on galaxies. 
\citet{Zhang2007} described a phase shift method to locate corotation radii in face-on barred and spiral galaxies, but this method assumes spirals and bars are quasi-steady modes in galactic disks.
Later, \citet{Buta2017} described a novel method for detecting resonances, attributing ``dark gaps" in galaxies to the location of corotation. 
However, if we accept that GALAKOS is a reliable model of a barred galaxy, then it provides an interesting alternative interpretation of the gaps. 
These same ``dark gaps" may indicate the location of the 4:1 ultra-harmonic resonance with the bar, in both ringed and non-ringed galaxies. 
This finding now opens up a new avenue to explore the effects of bar resonances on galaxies' structure using large samples of galaxy images. 

citet{Kim2016} analyzed the relation of the dark gaps with the bar properties for a sample of barred galaxies. They found that larger and stronger bars produced stronger dark gaps, indicating that the growth process of the bar is linked with the formation or the dark gaps. Similar to our cases, their dark gaps are located at radius smaller than the bar radius (see their fig. 5). In addition, they run a galaxy numerical model to understand the formation of the dark gaps. In this simulation the dark gaps formed were also located at a radius smaller than the bar radius. This is similar to what we obtain in our GALAKOS simulation. All these facts imply that if the dark gaps would be located at the corotation radius \citep{Buta2017}, all bars would be ultra-fast rotators. This would be in contradiction with the classical view of bars supported by the orbits of the x1-family \citep[e.g.][]{Contopoulos1980,Athanassoula1992}.

\begin{figure}[htb!]
\epsscale{1.15}
\centering
\hspace*{-.25cm}  
\plotone{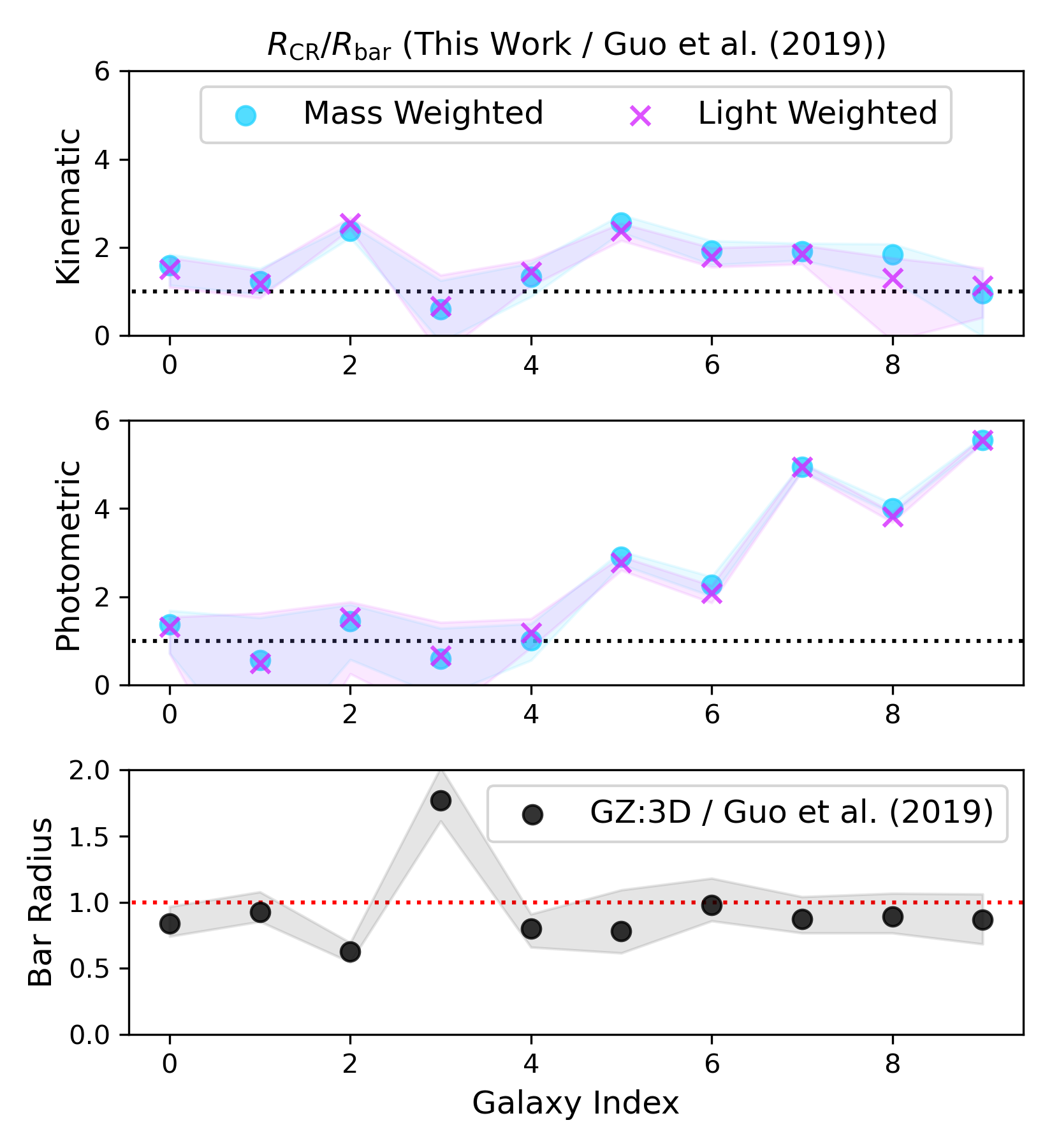}
\vspace*{-0.2cm}
\caption{The ratio of the estimated corotation radii and the estimates from the mass-weighted (cyan circle) and light-weighted (pink X) measurements from \citet{Guo2019} for galaxies overlapping in both samples. The top and middle panels are for the kinematic and photometric based estimates from \citet{Guo2019}, and shading encompasses both measurements' errors. {\it Bottom panel}: The ratio between our bar length estimate (Galaxy Zoo:3D) and the estimates in \citet[][units of arcsec]{Guo2019}. The ``Galaxy Index" is an arbitrary index describing each overlapping galaxy in our sample and that of \citet{Guo2019}.}
\label{fig:guo_compare}
\end{figure}

The most accurate method for measuring the bar pattern speed, which is used to derive the corotation radius, is using the kinematic and model-independent method proposed by \citet{Tremaine1984}. The method is, however, affected by multiple error sources, notably the determination of geometric parameters and modeling of the rotation curve \citep{Garma-Oehmichen2020}. 
Additionally, it is often best suited to work with early type galaxies.
\citet{Guo2019} used this method to measure the bar pattern speed and corotation radii of a sample of $53$ MaNGA galaxies. Of these, $10$ galaxies are also in our barred sample and have reliable estimates from our new method without considering any kinematic information.
Figure \ref{fig:guo_compare} shows the difference between our estimated corotation radius and the measurements from \citet{Guo2019} using their kinematic and photometric methods. 
While large discrepancies are present in some individual galaxies, the uncertainties combined with the kinematic method cannot rule out an overall agreement. 
The median value of the ratio between our estimate using $R_\mathrm{Ratio,CR}$ and the \citet{Guo2019} mass-weighted kinematic method is $\nicefrac{R_\mathrm{Cross,CR}}{R_\mathrm{kin,CR,m}} = 1.7^{+0.4}_{-0.2}$, showing that our method tends to estimate the corotation radius to be farther out than using the \citet{Tremaine1984} method. 
Figure \ref{fig:guo_compare} also shows differences between our bar length estimates from Galaxy Zoo:3D and the estimates derived in \citet{Guo2019}. In the future, applying the \citet{Tremaine1984} method to the full sample of MaNGA galaxies will allow for  more statistically robust validation of our method. 
The kinematic approach to identifying the corotation radius is likely still the most accurate method, but our new method explored in this work may be preferable for more face-on galaxies, whose kinematics are less reliable in projection and whose morphological features can be better characterized. 

Though the method used to measure bar length can impact estimates of resonance location, it is difficult to gauge which method is best. 
We note that the Galaxy Zoo:3D bar length estimates are generally consistent with estimates using other methods, including in our overlap with the work of \citet{Guo2019} with a median of $\nicefrac{R_\mathrm{bar,GZ:3D}}{R_\mathrm{bar,Guo}} = 0.9\pm0.1$. 
\citet{Guo2019} used a combination of three methods to estimate bar lengths, including the ellipticity radial profile, the ellipticity position angle profile, and the Fourier decomposition.
In \citet[][their Appendix]{Krishnarao2020b}, the Galaxy Zoo:3D bar lengths were tested more extensively with a Fourier-based method of \citet{Kraljic2012}, showing consistent results.

\begin{figure*}[htb!]
\centering
\plottwo{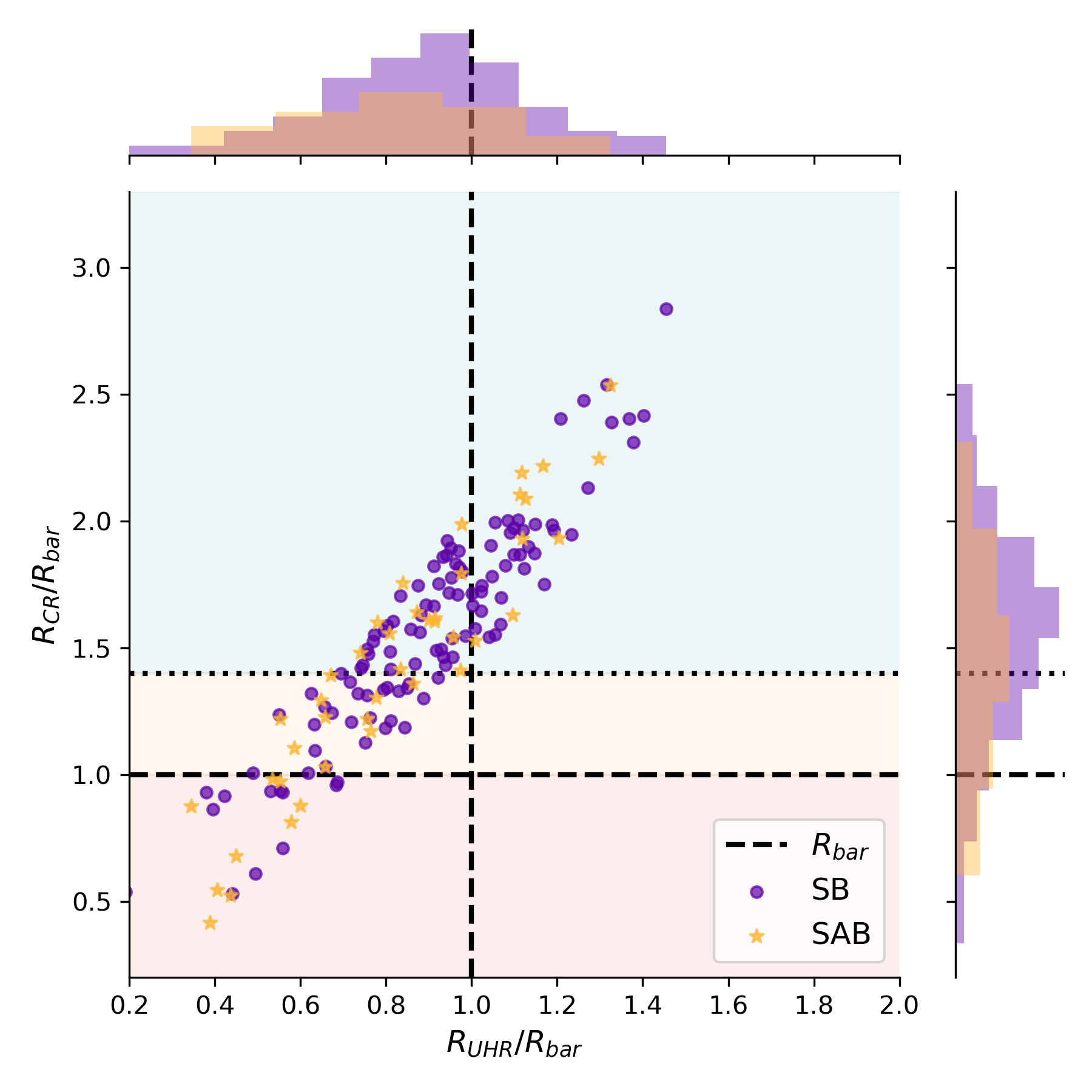}{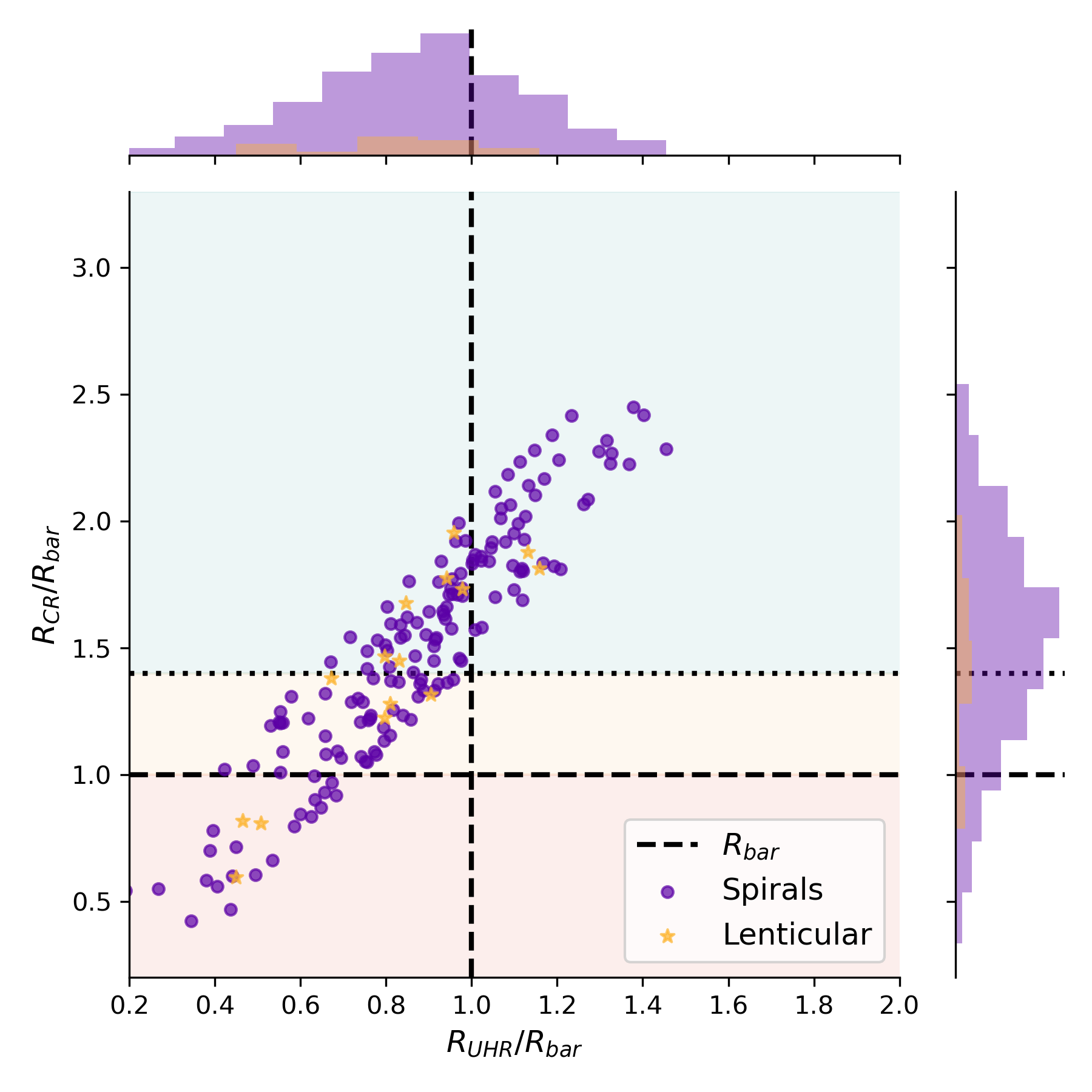}
\vspace*{-0.2cm}
\caption{Same as in Figure \ref{fig:obs_all_CR}, but separated by morphological types from Galaxy Zoo 2 \citep{Hart2016}, with SB vs SA (left panel; proxy for strong vs. weak bar) and Spirals vs. Lenticulars (middle panel). }
\label{fig:obs_all_CR_morph}
\end{figure*}

Also, we can explore the correlation between different galaxy morphological types and their bar speeds. 
Using the most likely morphological types as diagnosed in Galaxy Zoo 2 \citep{Hart2016}, we compare differences in the corotation radii. 
Figure \ref{fig:obs_all_CR_morph} shows the distribution of measured resonance radii separated by their morphological types. 
Several galaxies have a likely morphology from Galaxy Zoo 2 as SA, or not containing a bar. These galaxies likely have weaker bars than those with a likely morphology of SB, which would have bars more clearly visible in SDSS imaging. 
Here we use this difference as an approximate proxy for bar strength.
With our current sample, we can not robustly diagnose any correlation with morphology or bar strength and bar speeds. 
Understanding the impact of the bar strength on the bar speed could be studied in the future with more detailed analysis, but it is beyond the scope of this paper. 

In our analysis of individual galaxies, only $32\%$ of our full sample of $578$ galaxies showed a clear `dark-gap' signature in the PCA stellar mass surface density maps. While this is relatively low, it may still be possible to improve the success rate using deeper imaging data and to vary the amount of smoothing. In particular, it may be possible to identify ``dark-gaps" using the deep multi-band imaging from the DESI legacy imaging survey \citep{legacy_dr8}. This approach also does not involve the more complicated effects of spatial covariances resulting from dithered IFU observations. 

In this paper, we only consider the stellar mass and light concerning bar-driven resonances because the GALAKOS simulation lacks gas.
It has been shown that gas can significantly impact the stellar dynamics of bars, which our test simulations do not account for \citep[e.g.][]{Athanassoula2003,Combes2008}.
However, cosmological simulations with the same level of detail as GALAKOS that include gas are not yet available, but in the future will provide an ideal way to calibrate this gap method.
In addition, including simulations of ringed galaxies and other types will help determine the universality of the gap method on all barred galaxies.
Initial examination of the MaNGA emission line data suggests that the gas distribution may also closely track the azimuthal variations of the stellar density shown in Figure \ref{fig:obs_polar_map} based on a similar map using the \ha\ equivalent width. 
However, the precise location of the `dark-gaps' is slightly shifted from the stellar matter. 
\citet{Fraser-McKelvie2020} investigated the correlation between the \ha\ morphology of bars and their star formation rates, finding correspondence between different gas morphologies and the amount of quenching in bars.
To fully understand the nature of the gas with resonances, we must, in the future, similarly calibrate our metrics using hydrodynamic simulations.

\section{Conclusions}\label{sec:conclusion}

This paper has proposed a new method to measure the radius of both corotation and the ultra-harmonic resonance in galaxies without measurements of their internal kinematics. 
Our method is very similar to the `dark-gap' method of \citet{Buta2017}, but with a significant change in the interpretation, prompted by tests on the N-body simulation, GALAKOS \citep{D'Onghia2020}. 
We summarize our main findings below:
\begin{enumerate}[itemsep=-0mm,leftmargin=*]
    \item In the GALAKOS simulation, dark gaps are found in the stellar mass surface density, but appear not to be related with the $L_{4,5}$ Lagrangian points as was proposed by \citet{Buta2017}. Instead, the model favors the gaps to be linked to the inner 4:1 ultra-harmonic resonance.
    \item If the rotation of the galaxy hosting the `dark-gap' is known, then the corotation resonance can be precisely identified using Equation \ref{master_eq}.
    \item Alternatively, assuming a generally flat rotation curve, the radius of corotation can be predicted as $\nicefrac{R_{\mathrm{Ratio,CR}}}{R_{\mathrm{UHR}}} = 1.8\pm0.3$ but with decreased precision. 
    \item This method works best for galaxies with imaging, or IFU spaxels, at $\sim 0.2 - 0.4$ kpc scales, similar to MaNGA galaxies, so that the strong non-axisymmetric variations from the bar in the inner-most regions do not dominate.
    \item Applying this method to a sample of $578$ barred MaNGA galaxies reveals that on average, the MaNGA sample represents a population of bars that are slow rotators, but it can not be ruled out that the bars are fast rotators given the uncertainties, especially on the determination of the bar radius. 
    \item About $32\%$ of our barred galaxy sample show clear signatures of a `dark-gap' in the PCA-based resolved stellar mass surface density maps of \citet{Pace2019a,Pace2019b}.
    \item Our results for these individual galaxies are generally consistent with estimates derived using the \citet{Tremaine1984} method in previous work \citep{Guo2019}, though both methods have large systematic errors in our sample.
    \item None of the individual bars can be confidently determined to be slow rotators when considering their $2\sigma$ uncertainties.
\end{enumerate}

In particular, our method does not require kinematic information to determine the fundamental parameters describing the dynamics of bars. It can be used on large samples of barred galaxies from imaging surveys, such as the Legacy Survey \citep{legacy_dr8}. 
This method allows for the nature of rings and other structures in barred galaxies to be re-examined in terms of bar driven resonances, paving the way for better diagnostics of galaxies' internal dynamical structures. 
With the continued use of citizen science projects like Galaxy Zoo to build larger training samples, it may soon be possible to identify bar resonances with automated machine-vision techniques or neural networks.

\acknowledgments
DK and ZJP acknowledge support from the NSF CAREER Award AST-1554877. DK is supported by an NSF Astronomy and Astrophysics Postdoctoral Fellowship under award AST-2102490.
JALA is supported by the Spanish MINECO grant AYA2017-83204-P.

J.G.F-T gratefully acknowledges the grant support provided by Proyecto Fondecyt Iniciaci\'on No. 11220340, and also from ANID Concurso de Fomento a la Vinculaci\'on Internacional para Instituciones de Investigaci\'on Regionales (Modalidad corta duraci\'on) Proyecto No. FOVI210020, and from the Joint Committee ESO-Government of Chile 2021 (ORP 023/2021). 

Funding for the Sloan Digital Sky Survey IV has been provided by the Alfred P. Sloan Foundation, the U.S. Department of Energy Office of Science, and the Participating Institutions. SDSS acknowledges support and resources from the Center for High-Performance Computing at the University of Utah. The SDSS web site is \href{https://www.sdss.org}{www.sdss.org}.

SDSS is managed by the Astrophysical Research Consortium for the Participating Institutions of the SDSS Collaboration including the Brazilian Participation Group, the Carnegie Institution for Science, Carnegie Mellon University, the Chilean Participation Group, the French Participation Group, Harvard-Smithsonian Center for Astrophysics, Instituto de Astrof\'{i}sica de Canarias, The Johns Hopkins University, Kavli Institute for the Physics and Mathematics of the Universe (IPMU) / University of Tokyo, the Korean Participation Group, Lawrence Berkeley National Laboratory, Leibniz Institut f\"{u}r Astrophysik Potsdam (AIP), Max-Planck-Institut f\"{u}r Astronomie (MPIA Heidelberg), Max-Planck-Institut f\"{u}r Astrophysik (MPA Garching), Max-Planck-Institut f\"{u}r Extraterrestrische Physik (MPE), National Astronomical Observatories of China, New Mexico State University, New York University, University of Notre Dame, Observat\'{o}rio Nacional / MCTI, The Ohio State University, Pennsylvania State University, Shanghai Astronomical Observatory, United Kingdom Participation Group, Universidad Nacional Aut\'{o}noma de M\'{e}xico, University of Arizona, University of Colorado Boulder, University of Oxford, University of Portsmouth, University of Utah, University of Virginia, University of Washington, University of Wisconsin, Vanderbilt University, and Yale University.

\vspace{5mm}
\facilities{Sloan}

\software{astropy \citep{astropy,astropy2},
        matplotlib \citep{mpl}, 
        seaborn \citep{sns}, 
        scipy \citep{scipy}, 
        sdss-marvin \citep{Cherinka2019}, 
        bettermoments \citep{Teague2018}
          }



 \newcommand{\noop}[1]{}

\end{document}